\documentclass[preprint,onecolumn,nofootinbib]{revtex4}

\usepackage[colorlinks=true,linkcolor=blue,urlcolor=blue,filecolor=black,citecolor=red,pdfstartview=FitV,pdftitle={},pdfsubject={},pdfkeywords={},pdfpagemode=None,bookmarksopen=true]{hyperref}
\usepackage{graphicx}
\usepackage{amsmath}
\usepackage{amsfonts}
\usepackage{amssymb,ulem}
\usepackage{color}%
\usepackage{tikz,float}
\usepackage{dcolumn}
\usepackage{xcolor}

\begin{document}
\title{
	Entanglement Wedge Minimum Cross-Section in Holographic Axion Gravity Theories
}
\author{Fang-Jing Cheng $^{1}$}
\email{chengfj@stu2018.jnu.edu.cn}
\author{Zhe Yang $^{1}$}
\email{yzar55@stu2021.jnu.edu.cn}
\author{Chao Niu $^{1}$}
\email{niuchaophy@gmail.com}
\author{Cheng-Yong Zhang $^{1}$}
\email{zhangcy@email.jnu.edu.cn}
\author{Peng Liu $^{1}$}
\email{phylp@email.jnu.edu.cn}
\thanks{corresponding author}
\affiliation{
	$^1$ Department of Physics and Siyuan Laboratory, Jinan University, Guangzhou 510632, China
}

\begin{abstract}

	We study the mixed state entanglement properties in two holographic axion models by examining the behavior of the entanglement wedge minimum cross section (EWCS), and comparing it with the holographic entanglement entropy (HEE) and mutual information (MI). We find that the behavior of HEE, MI and EWCS with Hawking temperature is monotonic, while the behavior with the axion parameter $k$ is more rich, which depends on the size of the configuration and the values of the other two parameters. Interestingly, the EWCS monotonically increases with the coupling constant $\kappa$ between the axion field and the Maxwell field, while HEE and MI can be non-monotonic. It suggests that the EWCS, as a mixed state entanglement measure, captures distinct degrees of freedom from the HEE and MI indeed. We also provide analytical understandings for most of the numerical results.

\end{abstract}
\maketitle
\tableofcontents

\section{Introduction}
\label{sec:introduction}

In recent years, research on quantum entanglement has attracted more and more attention across many different fields, including quantum information, condensed matter physics and quantum gravity \cite{Donos:2012js,Ryu:2006bv,Susskind:2014rva,Dong:2016fnf}. Many measures of quantum entanglement have been found to diagnose the quantum phase transition of strongly correlated systems, which opens a gate to the study of condensed matter physics from an information perspective. Furthermore, quantum entanglement is also considered crucial in the emergence of spacetime, which brings a new understanding for the essence of spacetime \cite{Osterloh:2002na,Amico:2007ag,Wen:2006topo,Kitaev:2006topo,Hubeny:2007xt,Lewkowycz:2013nqa,Dong:2016hjy}.

There are many measures of quantum entanglement, such as entanglement entropy (EE), mutual information (MI) and entanglement of purification (EoP). All of them can describe quantum entanglement to certain extent. The EE is only suitable for describing pure state entanglement. But most quantum systems are in mixed state in nature, so it is necessary to describe them with other measures, such as MI and EoP, which can extract the quantum as well as classical correlations from the mixed state systems. However, the entanglement measures are notoriously difficult to calculate, which has become the main obstacle in the study of quantum entanglement \cite{Terhal:2002,Horodecki:2009review}.

Over the past two decades, holographic duality theory has been widely adopted in many fields of physics \cite{tHooft:1993dmi,Susskind:1994vu,Maldacena:1997re,Hartnoll:2014lpa}. Recently, the measures of quantum entanglement of strongly correlated systems have been considered to be related to the geometric quantity of dual gravity systems, which provides a simple geometric method for calculating the quantum entanglement. It was first proposed that the holographic entanglement entropy (HEE), which is the holographic version of EE, is proportional to the area of the minimum surface in the dual gravity system \cite{Ryu:2006bv}, and has many applications in diagnosing the phase transitions \cite{Nishioka:2006gr,Klebanov:2007ws,Pakman:2008ui,Zhang:2016rcm,Zeng:2016fsb,Ling:2015dma,Ling:2016wyr}. After that, more holographic duality relations between the measures of quantum entanglement and the geometric quantity of dual gravity system have been proposed. For example, holographic R\'enyi entropy is proportional to the minimum area of cosmic branes \cite{Dong:2016fnf}. Moreover, the butterfly effects and quantum complexity and many other measures have been proposed and studied in \cite{Shenker:2013pqa,Sekino:2008he,Maldacena:2015waa,Blake:2016wvh,Blake:2016sud,Ling:2016ibq,Ling:2016wuy,Wu:2017mdl,Liu:2019npm,Brown:2015lvg,Brown:2015bva,Chapman:2016hwi,Ling:2018xpc,Chen:2018mcc,Yang:2019gce,Ling:2019ien,Kudler-Flam:2020url,BabaeiVelni:2020wfl,Sahraei:2021wqn,Khoeini-Moghaddam:2020ymm,KumarBasak:2020eia}. Entanglement of purification (EoP) and negativity has been associated with the minimum area of entanglement wedge minimum cross section (EWCS) in the dual gravity spacetime \cite{Takayanagi:2017knl,Tamaoka:2018ned}. Recently, our work proposed that EWCS can describe the thermal phase transitions of mixed state systems \cite{Liu:2021rks,Liu:2019qje,Huang:2019zph,Liu:2020blk}.

At present, HEE, as the earliest progress in the subject of holographic quantum information, has been widely studied. However, the studies on entanglement of mixed states such as MI and EWCS are still far from comprehensive. Holographic linear axion models are simple enough to have analytical solutions while having rich physical content, such as momentum dissipation, phonons, etc \cite{Baggioli:2014roa}. Therefore, in this paper, we will study the properties of mixed state entanglement in linear axion models in detail.

We organize this paper as follows: in Sect. \ref{sec:Information_related_quantities}, we review the concepts of three entanglement measures (HEE, MI, EWCS) and the configurations we will study. In Sect. \ref{Model_1}, we discuss the properties of EWCS in the first linear axion model. In Sect. \ref{Model_2}, we discuss the properties of three entanglement measures (HEE, MI, EWCS) in the second linear axion model. Finally, we summarize in Sect. \ref{sec:discuss}.

\section{Information related quantities}
\label{sec:Information_related_quantities}
In this section, we introduce the concepts and calculation methods of HEE, MI and EWCS.

\subsection{Holographic entanglement entropy}
\label{sec:HEE}

\begin{figure}
	\begin{tikzpicture}[scale=1]
		\node [above right] at (0,0) {\includegraphics[width=7.5cm]{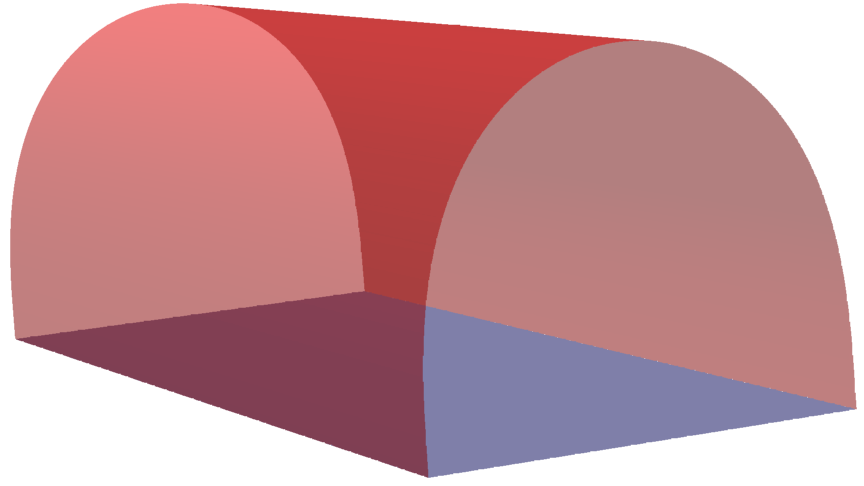}};
		\draw [right,->,thick] (3.85, 0.22) -- (6.25, 0.58) node[above] {$x$};
		\draw [right,->,thick] (3.85, 0.22) -- (1.25, 1.08) node[below] {$y$};
		\draw [right,->,thick] (3.85, 0.22) -- (3.7, 3.125) node[above] {$z$};
		\draw [right,-,thick,black] (3.85, 0.22) -- (4.53, 0.038);
		\draw [right,-,thick,black] (7.56, 0.79) -- (8.25, 0.633);
		\draw [right,-,thick,black] (4.195, 0.124) -- node[below] {$w$} (7.895, 0.7115) ;
	\end{tikzpicture}
	\begin{tikzpicture}[scale=1]
		\node [above right] at (0,0) {\includegraphics[width=7.5cm]{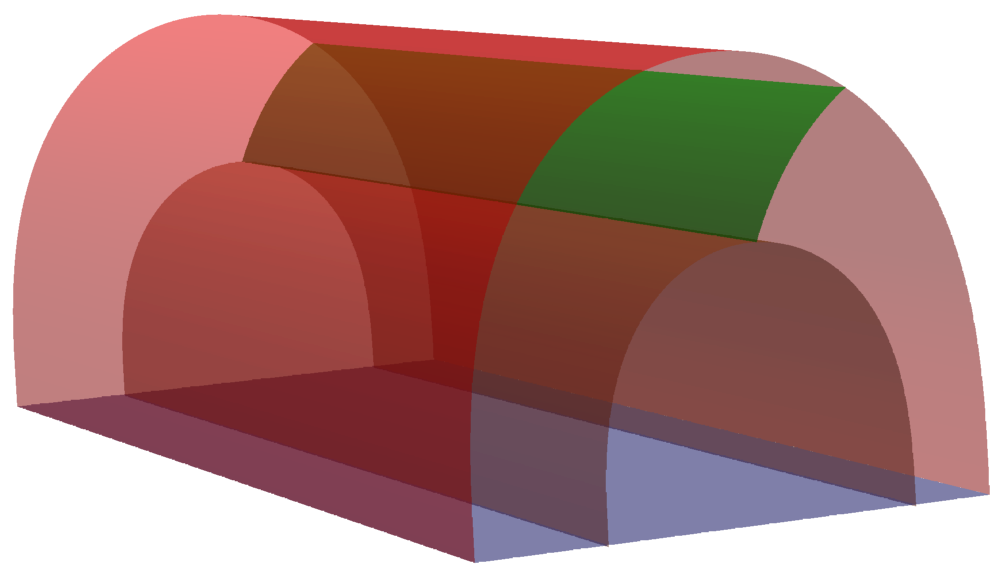}};
		\draw [right,->,thick] (3.67, 0.22) -- (6.25, 0.55) node[above] {$x$};
		\draw [right,->,thick] (3.67, 0.22) -- (1.25, 1.05) node[below] {$y$};
		\draw [right,->,thick] (3.67, 0.22) -- (3.6, 3.125) node[above] {$z$};
		\draw [right,-,thick,black] (3.67, 0.22) -- (4.35, 0.03);
		\draw [right,-,thick,black] (4.6004, 0.3454) -- (5.3004, 0.1554);
		\draw [right,-,thick,black] (6.9224, 0.6424) -- (7.65789, 0.473254);
		\draw [right,-,thick,black] (7.46, 0.71) -- (8.15, 0.55);
		\draw [right,-,thick,black] (4.01, 0.125) --  node[shift={(2.5*22.5:-5pt)}] {$a$} (4.9504, 0.2504);
		\draw [right,-,thick,black] (4.9504, 0.2504) --  node[shift={(6*22.5:-5pt)}] {$b$} (7.26515, 0.557827);
		\draw [right,-,thick,black] (7.26515, 0.557827) --  node[shift={(5*22.5:-2pt)}] {$c$} (7.805, 0.63);
	\end{tikzpicture}
	\caption{
		A cartoon of the studied configuration. (a) The upper figure shows the configuration used to discuss HEE. (b) The lower figure shows the configuration used to discuss MI and EWCS.
	}
	\label{cartoon1}
\end{figure}
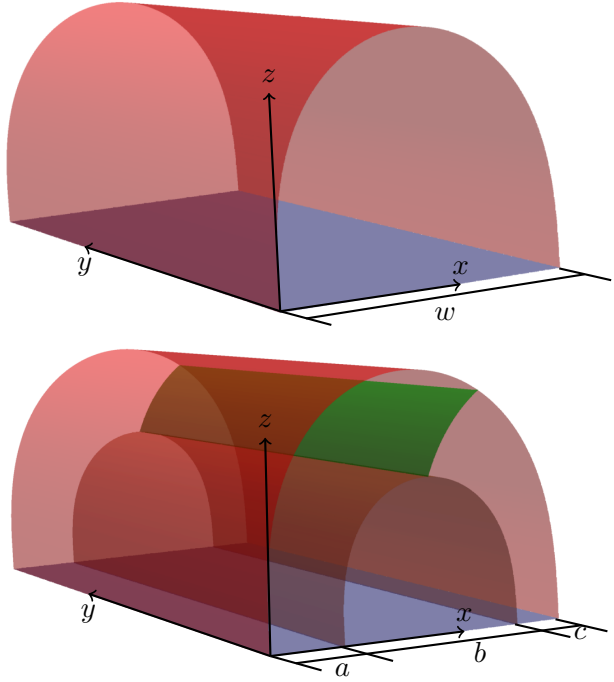

HEE has been considered proportional to the area of the minimum surface extending into the bulk \cite{Ryu:2006bv}. Although this duality is very concise, it is difficult to calculate. To simplify the calculation, we consider the infinite strip configuration along $y$-axis (see Fig. \ref{cartoon1} (a)) in a homogeneous spacetime,
\begin{equation}
	d s ^ { 2 } = g _ { t t } d t ^ { 2 } + g _ { x x }  d x ^ { 2 } + g _ { y y } d y ^ {2} + g _ { z z } d z ^ {2},
\end{equation}
where $z = \frac{r_h}{r}$, which means that $z = 0$ is the AdS boundary and $z = 1$ the horizon of the black brane. This configuration is invariant along the $y$-axis, so its width $w$ is enough to characterize it. The minimum surface can be parameterized as $z(x)$ on the $x-z$ plane. The area of the minimum surface is,
\begin{equation}
	A = L_y \int \sqrt {g _ { y y } ( g _ { x x } dx ^ {2} + g _ { z z } dz ^ {2}) } = L_y \int \sqrt {g _ { y y } ( g _ { x x } + g _ { z z } z'(x) ^ {2} )} dx = L_y \tilde A \varpropto \tilde A ,
\end{equation}
where $\tilde A = \int \sqrt {g _ { y y } ( g _ { x x } + g _ { z z } z'(x) ^ {2} )} dx$. $L_y = \int dy$ is the infinite length of the $y-$axis, which we ignore since it is a constant. Therefore, we only focus on $\tilde A$. Variating $\tilde A$ leads to the equations of motion of the minimum surface. Due to the divergence of asymptotic AdS at $z = 0$, the parameterization $z(x)$ is not suitable for solving the minimum surface. A better candidate parameterization is the angle $\theta$,
\begin{equation}
	\tilde A = \int \sqrt {g _ { y y }(z(\theta)) ( g _ { x x }(z(\theta)) x'(\theta) ^ {2} + g _ { z z }(z(\theta)) z'(\theta) ^ {2})} d \theta,
\end{equation}
with $\tan\theta = \frac{z}{x}$ \cite{Liu:2019qje,Huang:2019zph}. Also, the equations of motion for $z(\theta)$ and $x(\theta)$ can be obtained by variating $\tilde A$. Asymptotic AdS will lead to a divergent $\tilde A$. Therefore, in practical calculations, we need to subtract a common divergent term $\tilde A_{AdS}$ brought by the AdS. We treat the remaining finite term $S$ as the EE,
\begin{equation}\label{eq:sdef}
	S = \tilde A - \tilde A_{AdS}.
\end{equation}
When the width of the configuration is very small, the minimum surface resides near the AdS boundary. Therefore, the HEE with small configuration is the same as that of AdS. However, when the configuration is large (i.e. the width $w$ of the infinite strip is large), the minimum surface will be close to the horizon of the black brane. At this time, the HEE will be mainly contributed by the thermal entropy. The HEE will behave as
\begin{equation}
	S \simeq s w,
\end{equation}
where $s$ is the thermal entropy density. In other words, HEE with large configurations is dictated by the thermal entropy, which may bury the real quantum entanglement. This is the understanding of the influence of thermal entropy contained in HEE from the perspective of holography \cite{Liu:2019qje}.

\subsection{Mutual information}

The mutual information between two disjoint subsystems $A$ and $C$ separated by $B$ (with widths $a, \, b$ and $c$) is defined as
\begin{equation}
	I( A ; C )= S ( A )+ S ( C ) - S ( A \cup C ),
\end{equation}
where $S(\#)$ is the entanglement entropy of the region $\#$. More specifically, the minimum surface for $A\cup C$ has two locally minimum candidates from which we choose the minimum as the right one,
\begin{equation}
	S ( A \cup C ) = \min [ S ( A ) + S ( C ) , S ( B ) + S ( A + B + C)].
\end{equation}
When $S(A \cup C) = S (A) + S (C)$, we have $I (A; C) = 0$. In this case, there is no entanglement between $A$ and $C$ systems, and MI is trivial. When $S (A \cup C) = S (B) + S (A + B + C)$, we have $I (A ; C) > 0$. In this case, MI is nontrivial \cite{Fischler:2012uv}.

Obviously, the definition of MI is directly based on the definition of HEE. However, MI by definition cancels out the divergent term of HEE. In addition, MI partially offsets the influence of thermodynamic entropy \cite{Fischler:2012uv}. This shows that MI is more suitable than HEE to measure the entanglement of mixed states.

\subsection{Entanglement Wedge Minimum Cross Section}

In addition to calculating the entanglement of the mixed state by directly subtracting the influence of the thermal entropy like MI, we can also purify the mixed state by introducing additional degrees of freedom, so as to extract the quantum entanglement. This is called the entanglement of purification (EoP) \cite{Terhal:2002}. Specifically, the EoP is defined as the minimum entanglement entropy among all possible purification. Later, it was proposed that EoP is proportional to the minimum cross section in the entanglement wedge of the dual spacetime \cite{Takayanagi:2017knl} (see the green section in Fig. \ref{cartoon1} (b)),
\begin{equation}
	E _ {W} ( \rho _ { A B }) = \underset{ \Sigma _ { A B }}{ \min } \bigg( \frac{ Area ( \Sigma _ { A B })}{ 4 G _ { N }} \bigg).
\end{equation}
In addition to EoP, the reflected entropy, negativity, etc \cite{Chaturvedi:2016rcn,Dutta:2019gen}, are also proposed dual to EWCS. From the definition, EWCS is neither dominated by the AdS boundary nor dominated by the near horizon geometry. Therefore, EWCS is not dictated by thermal entropy, and hence is free from the divergence problem caused by AdS.

The EWCS with symmetrical configuration $(a=c)$ is obviously a vertical plane, so its calculation is very simple. However, it is very difficult to calculate EWCS for asymmetric configurations $(a\neq c)$. In \cite{Liu:2019qje}, we proposed a numerical algorithm to calculate EWCS using translation invariance. Later in \cite{Liu:2020blk}, we proposed an improved algorithm based on Newton iteration, which greatly improves the accuracy of EWCS calculation.

\subsection{Classification of configuration}

To more clearly discuss the relationship between configuration and information related physical quantities, we divide configuration into three categories: large, intermediate and small configurations.

For HEE, we categorize configuration according to the behavior under the limit of width: if the behavior of HEE with parameters is the same as the case when the configuration width tends to zero, the corresponding configuration is small; if the behavior of HEE with parameters is the same as the case when the configuration width tends to infinity, the corresponding configuration is large; otherwise, the corresponding configuration is intermediate. The advantage of this classification is that the small configuration directly reflects the properties of the AdS boundary, and the large configuration directly reflects the properties of the thermal entropy. In addition, the behavior of HEE with parameters under intermediate configuration may be very complicated and need extra understandings.

For MI and EWCS, the configurations here involve the widths of $A$, $B$ and $C$, so the classification of large, intermediate and small configurations in HEE case cannot be straightforwardly applied.
We adopt the following conditions: if the minimum surface corresponding to $A\cup B\cup C$ is very close to the AdS boundary, then the corresponding configuration is small; if the minimum surface corresponding to $A$ and $C$ are all very close to the horizon of the black brane, then the corresponding configuration is large. In other cases, the corresponding configuration is intermediate.

Next, we discuss HEE, MI and EWCS on two different axion models.

\section{Model I}
\label{Model_1}

\subsection{Background solution}

In the previous work \cite{Huang:2019zph}, HEE, MI and EWCS of symmetric configuration in model I are discussed. This section will supplement the discussion of EWCS in model I with asymmetric configurations. The action of the model I is,
\begin{equation}
	S = \int d ^ {4} x \sqrt { - g } \bigg[ R + 6 - V(X) - \frac{1}{4} F _ { \mu \nu } F ^ { \mu \nu }\bigg],
\end{equation}
where $V(X)$ is the kinematic term of the axion fields $\psi^I$. Ansatz and the corresponding solutions are,
\begin{equation}\label{eq:ansatz}
	\begin{aligned}
		 & ds ^ {2} = - f(r) dt ^ {2} + \frac{ 1 }{ f(r) } dr ^ {2} + r ^ {2} \delta _ { i j } dx ^ {i} dx ^ {j}, \\ & A _ {t} = \mu - \frac{\rho}{r} , \quad \psi ^ { I }= k \delta ^ {I} _ {i} x ^ {i}, \\ & X = \frac{1}{2} g ^ { a b } \delta _ { I J } \partial _ {a} \psi ^ {I} \partial _ {b} \psi ^ {J}, \quad \text{where} \; i,j,I,J = 1,2,
	\end{aligned}
\end{equation}
with
\begin{equation}
	\ \ \ f(r) = \frac{\rho ^ {2}}{ 4 r }\left(\frac{1}{r} - \frac{1}{r _ {h}}\right) + \frac{ 1 }{ 2 r } \int _ { r _ {h}} ^ { r } \bigg[ 6 u ^ {2} - V\left(\frac{ k ^ {2}}{ u ^ {2}}\right) u ^ {2} \bigg] du.
\end{equation}
The $\psi^I$ represents the linear axion field with a constant linear factor $k$. The vector field $A$ is the Maxwell field, and $\mu$, $\rho$ represent the chemical potential and the charge density of the dual field theory. For regularity, the value of the Maxwell field on the horizon vanishes, thus we have $\rho = \mu \, r_h$.

For simplicity, we choose $V(X) = X^2$. The system \eqref{eq:ansatz} is invariant under the following rescaling,
\begin{equation}\label{eq:scalingnew}
	\begin{aligned}
		 & (t \, ,\, x  \, ,\, y) \rightarrow \alpha(t  \, , \,x  \, ,\, y) ,      \\
		 & (r  \, ,\, k  \, ,\, \mu) \rightarrow (r  \, ,\, k  \, ,\, \mu)/\alpha, \\
		 & (f(r)  \, , \,\rho) \rightarrow (f(r)  \, ,\, \rho)/ \alpha ^ {2}.
	\end{aligned}
\end{equation}
We focus on scaling-invariant physical quantities, so we adopt $\mu$ as the scaling unit by setting $\mu$ = 1. Hawking temperature is given by,
\begin{equation}
	T = \frac{f'(r)}{ 4 \pi} \bigg|_{r = r _ {h}}= - \frac{r _ {h} ^ {2} + 2 k ^ {4} - 12 r _ {h} ^ {4}}{16 \pi r _ {h} ^ {3}}.
\end{equation}

\subsection{Entanglement Wedge Minimum Cross Section}

\begin{figure}
	\centering
	\includegraphics[width = 0.6\textwidth]{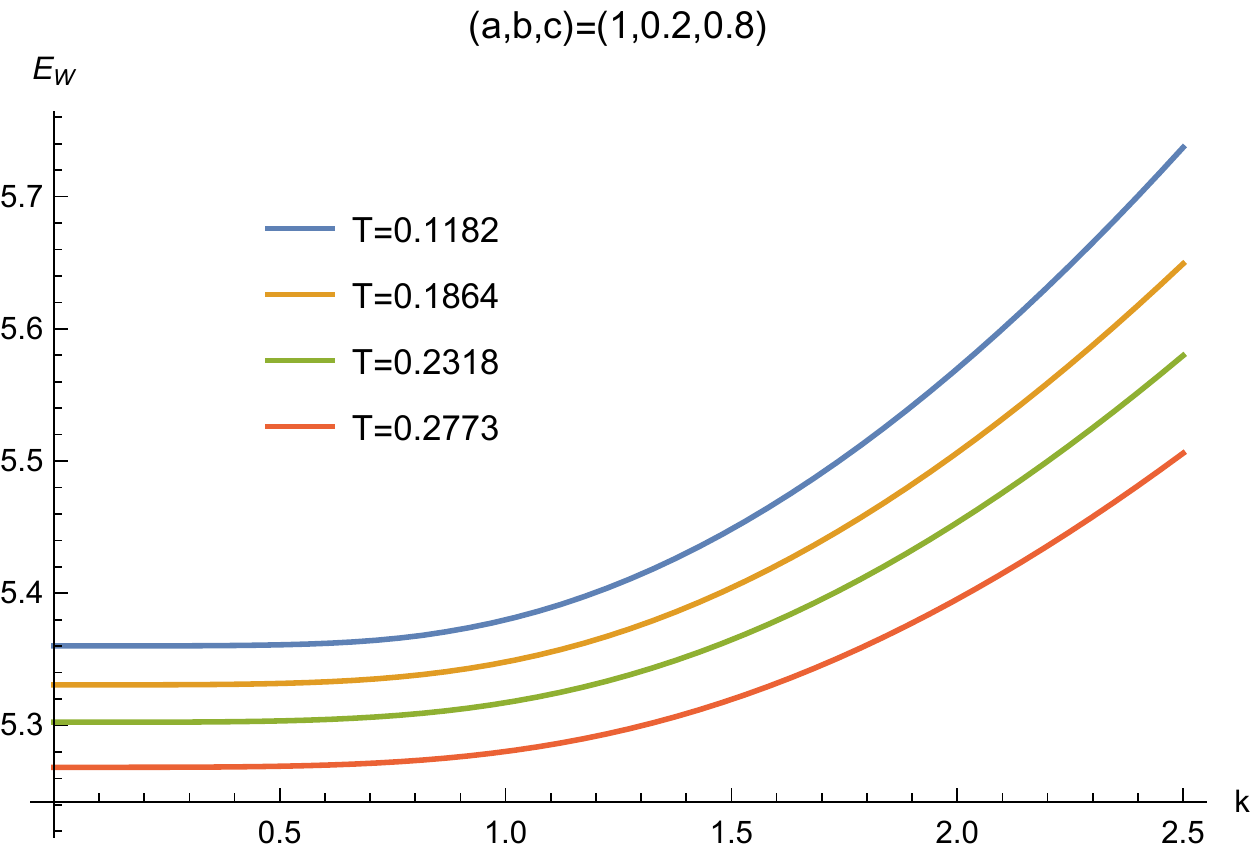}
	\caption{
		The relation between EWCS and $k$ in the large configuration for model I. It can be seen from the figure that EWCS increases monotonically with $k$, and it also has the same monotonic behavior under other large configurations and $T$.
	}
	\label{fig:case1_large_configuration_EoPk1}
\end{figure}

We find that the behavior of EWCS with $k$ depends on the details of the configurations and the value of $T$. In large configurations, the EWCS increases monotonically with $k$, independent of the value of $T$. We show this phenomenon in Fig. \ref{fig:case1_large_configuration_EoPk1}.
Moreover, when $k$ is small, the slope is very small.

\begin{figure}
	\centering
	\includegraphics[width = 0.6\textwidth]{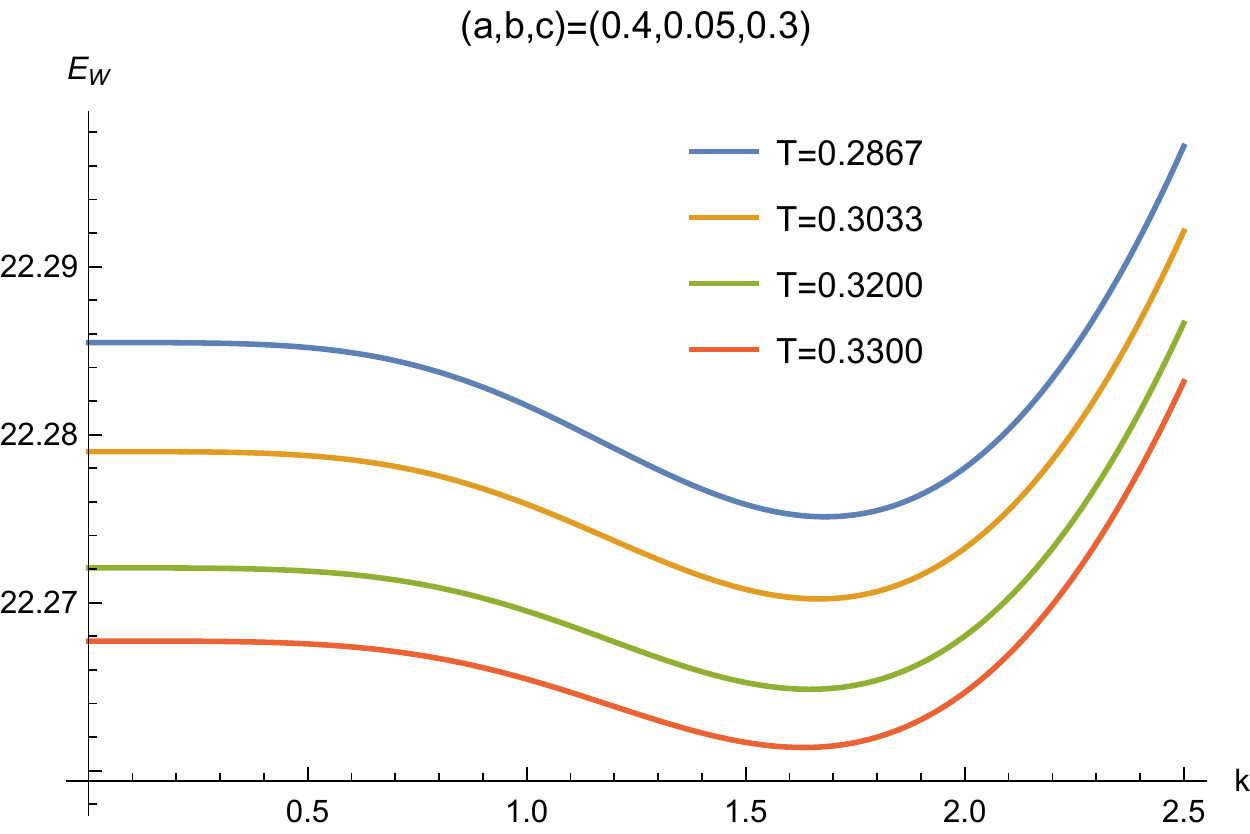}
	\caption{
		The relation between EWCS and $k$ in the small configuration for model I. It can be seen that when $T$ is appropriate, EWCS first decreases and then increases with $k$.
	}
	\label{fig:case1_small_configuration_EoPk1}
\end{figure}

\begin{figure}
	\centering
	\includegraphics[width = 0.49\textwidth]{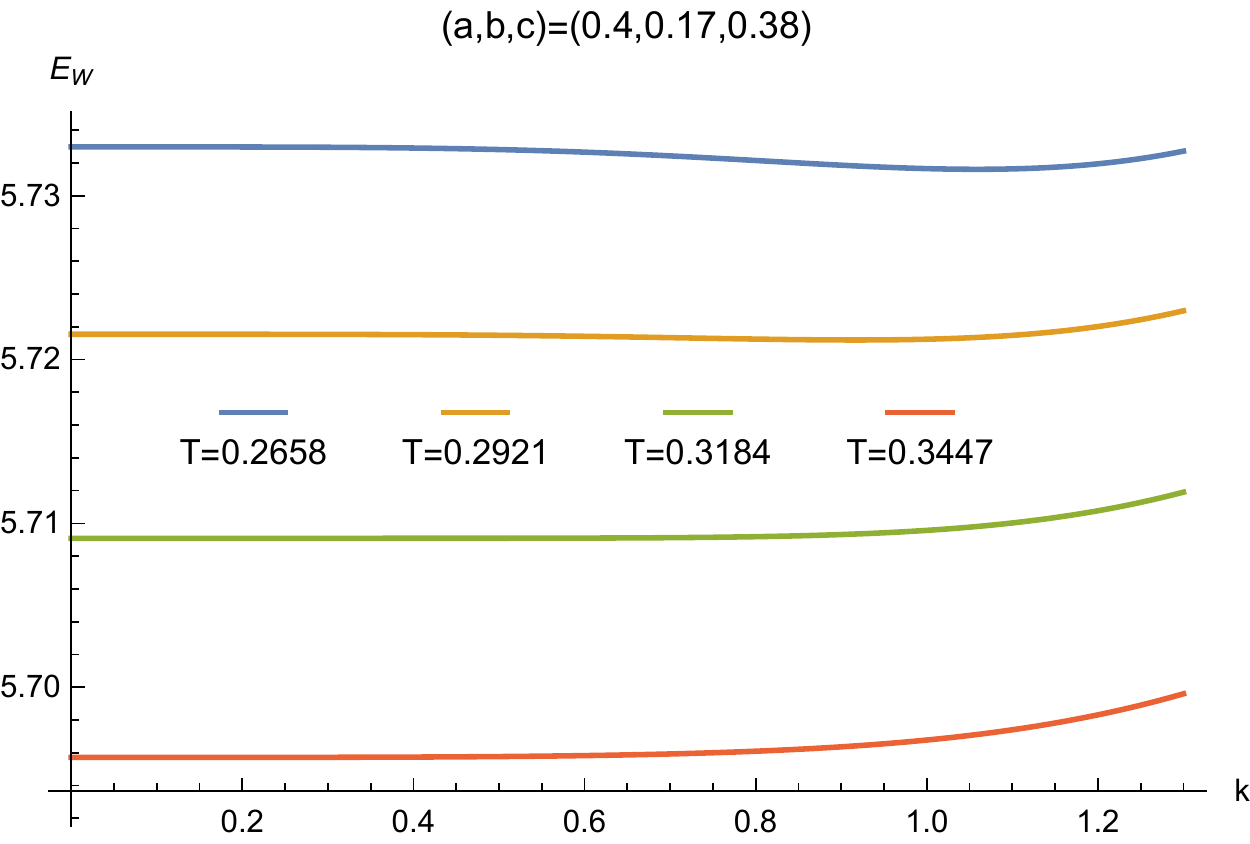}
	\includegraphics[width = 0.49\textwidth]{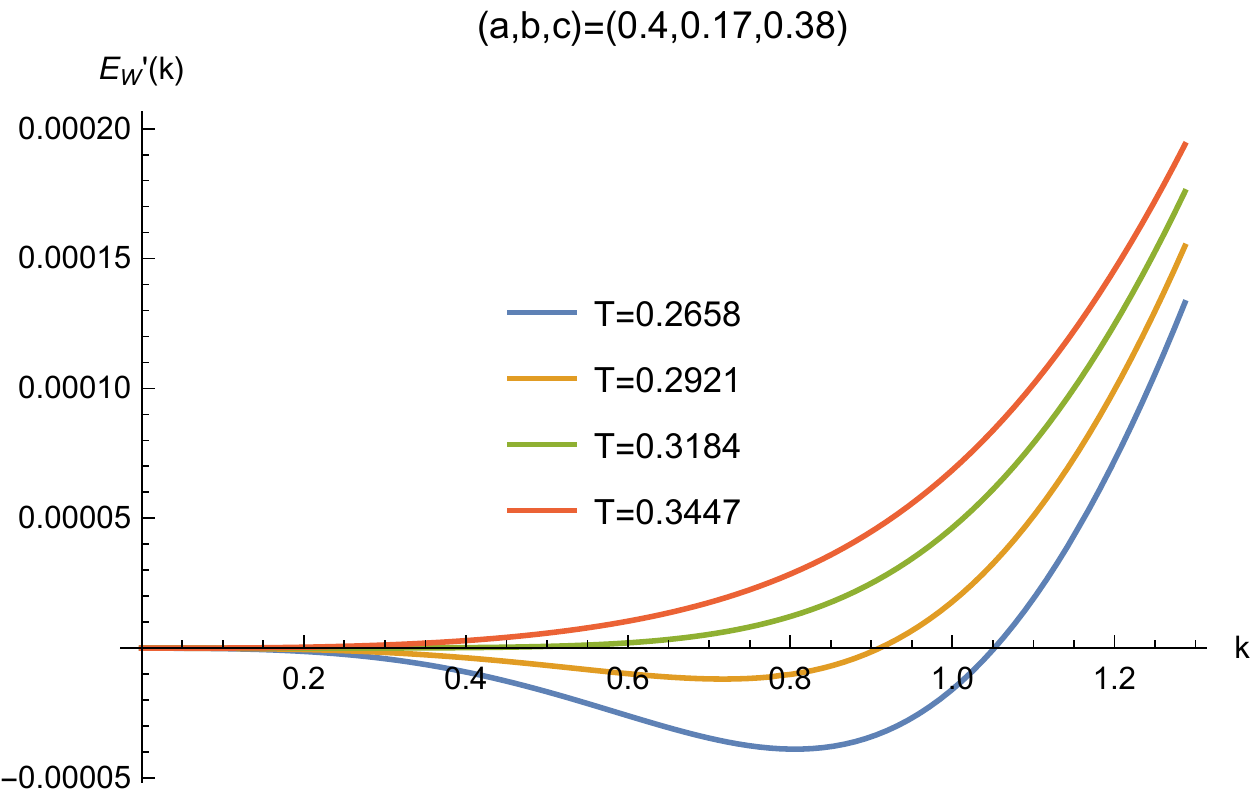}
	\caption{
		The figure on the left shows the relation between EWCS and $k$ in another small configuration for model I. The figure on the right shows the relation between the derivative of EWCS to $k$ and $k$ for model I. It can be seen from the figure on the left that the decreasing behavior of EWCS with $k$ weakens or even disappears when the temperature increases. This conclusion can be seen more clearly in the figure on the right. Moreover, we can see from the right figure that when EWCS is non-monotonic with $k$, the increase of $T$ will make the minimum point shift to the left.
	}
	\label{fig:case1_small_configuration_EoPk2}
\end{figure}

In small configurations, when the value of $T$ is appropriate, the non-monotonic behavior of EWCS with $k$ will appear, as shown in Fig. \ref{fig:case1_small_configuration_EoPk1}.
Specifically, the EWCS decreases first and then increases with $k$. We also find that the minimum point will shift to the left with the increase of temperature, and the decreasing behavior of EWCS with $k$ will weaken or even disappear when the configuration is fixed. These conclusions can be more easily drawn from Fig. \ref{fig:case1_small_configuration_EoPk2}.

\begin{figure}
	\centering
	\includegraphics[width = 0.6\textwidth]{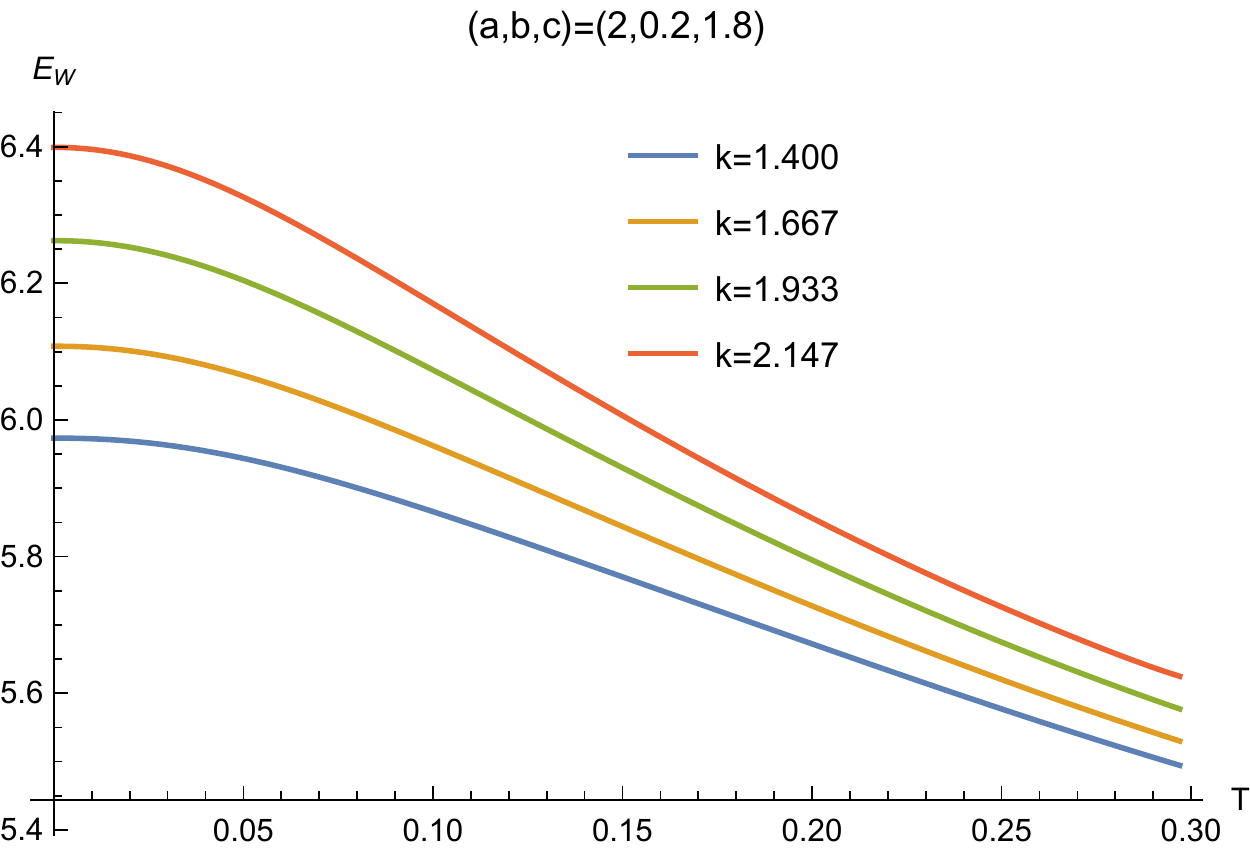}
	\caption{The relation between EWCS and $T$ for model I. It can be seen from the figure that EWCS decreases monotonically with $T$, and it also has the same monotonic behavior under other configurations and parameters.}
	\label{fig:case1_EoPT1}
\end{figure}

Moreover, we find that EWCS decreases monotonically with $T$, but is independent of configuration and $k$ (see Fig. \ref{fig:case1_EoPT1}).
The behavior that EWCS decreases with $T$ can be understood from the perspective of dual physics. Increasing the temperature of the quantum system can destroy the entanglement between subregions, so the physical quantities related to the entanglement of mixed states should decrease with the increase of temperature.

Next, we discuss the relationship between the EWCS of asymmetric configuration discussed here and the EWCS of symmetric configuration have discussed in \cite{Huang:2019zph}. First, the relationship between EWCS and temperature is the same in both cases, which is also in line with the physical expectation that high temperature breaks the entanglement. However, the EWCS with asymmetric configuration can exhibit non-monotonic behavior with $k$. This shows that asymmetric configurations can capture more information than symmetric configurations, which is also one of our main motives to study asymmetric configuration.

\section{Model II}
\label{Model_2}

\subsection{Background solution}
The action for model II is \cite{Baggioli:2014roa,Alberte:2015isw,Gouteraux:2016wxj},
\begin{equation}
	S = \int d ^ {4} x \sqrt { - g } \bigg[ R + 6 - V(X) - \frac{1}{4} ( 1 + \kappa \text{Tr}[X]) F _ { \mu \nu } F ^ { \mu \nu }\bigg],
\end{equation}
where $V(X)$ is the kinematic term of the axion fields, $\kappa$ represents the coupling constant between the axion and the Maxwell field. The ansatz and the background solutions are,
\begin{align}
	ds ^ {2} & = - f(r) dt ^ {2} + \frac{ 1 }{ f(r) } dr ^ {2} + r ^ {2} \delta _ { i j } dx ^ {i} dx ^ {j}, \\  A _ {t} &= \mu - \frac{ \pi \rho }{ 2 k \sqrt{ | \kappa | }} + \rho \frac{ _ {2} F _ {1}( 1 , -\frac{1}{2} ; \frac{1}{2} ; - \frac{ r ^ {2} }{ \kappa k ^ {2} } ) -1}{ r }, \quad \psi ^ { I }= k \delta ^ {I} _ {i} x ^ {i}, \\  X &= \frac{1}{2} g ^ { a b } \delta _ { I J } \partial _ {a} \psi ^ {I} \partial _ {b} \psi ^ {J}, \quad \text{where} \; i,j,I,J = 1,2,
\end{align}
with
\begin{equation}
	f(r) = \frac{ 1 }{ 2 r } \int _ { r _ {h}} ^ { r } \bigg[ \bigg( 6 - V\left( \frac{ k ^ { 2 }}{ s ^ { 2 }}\right) \bigg) s ^ { 2 } - \frac{ 1 }{ 2 } \frac{ \rho ^ { 2 }}{ s ^ { 2 } + \kappa k ^ { 2 }} \bigg] ds.
\end{equation}
The $\psi ^ I$ represents the linear axion field with a constant linear factor $k$. Like model I, The vector field $A$ is the Maxwell field, and $\mu$, $\rho$ represent the chemical potential and the charge density of the dual field theory. The coupling $\kappa \text{Tr}[X] F_{ \mu \nu } F ^ { \mu \nu}/4$ between the axion field and the Maxwell is the main difference between the present model II and model I. When $\kappa = 0$, model II will return to model I. Therefore, we will focus on the influence of this coupling term on the physical quantities related to quantum information.

Again, we specifically choose $V (X) = X^2$ for concreteness. The regularity condition for the Maxwell field on the horizon is,
\begin{equation}\label{eq:horizoncondition}
	\mu - \frac{ \pi \rho }{ 2 k \sqrt{ | \kappa | }} + \rho \frac{ _ {2} F _ {1}( 1 , -\frac{1}{2} ; \frac{1}{2} ; - \frac{ r _ {h} ^ {2} }{ \kappa k ^ {2} } )-1}{ r _ {h} } =0.
\end{equation}
The setup of this model is invariant under the following rescaling,
\begin{equation}\label{eq:invariants}
	\begin{aligned}
		(t \, ,\, x  \, ,\, y) \rightarrow    & \alpha(t  \, , \,x  \, ,\, y) ,    \\
		(r  \, ,\, k  \, ,\, \mu) \rightarrow & (r  \, ,\, k  \, ,\, \mu)/\alpha,  \\
		(f(r)  \, , \,\rho) \rightarrow       & (f(r)  \, ,\, \rho)/ \alpha ^ {2}.
	\end{aligned}
\end{equation}
We focus on scaling-invariant physical quantities, so again we adopt $\mu$ as the scaling unit by setting $\mu$ = 1. Hawking temperature is given by,
\begin{equation}\label{eq:hawkingtemperature2}
	T = \frac{ 1 }{ 16 \pi } \bigg[ 2 r _ { h } \bigg( 6 - V \bigg( \frac{ k ^ {2}}{ r _ { h } ^ { 2 }} \bigg) \bigg) - \frac{ \rho ^ { 2 }}{ r _ { h } ^ { 3 } ( 1 + \kappa ( k ^ { 2 } / r _ { h } ^ { 2 }))} \bigg].
\end{equation}
In addition, we find that when $\kappa < 0$, the black brane solution can be complex in certain regions of the bulk. Therefore, throughout this paper only we discusses the case of $\kappa\geqslant 0$.

\subsection{Holographic entanglement entropy}

In this subsection, we explore the relationship between HEE and system parameters $(k, T, \kappa)$ in model II.

\begin{figure}
	\centering
	\includegraphics[width = 0.6\textwidth]{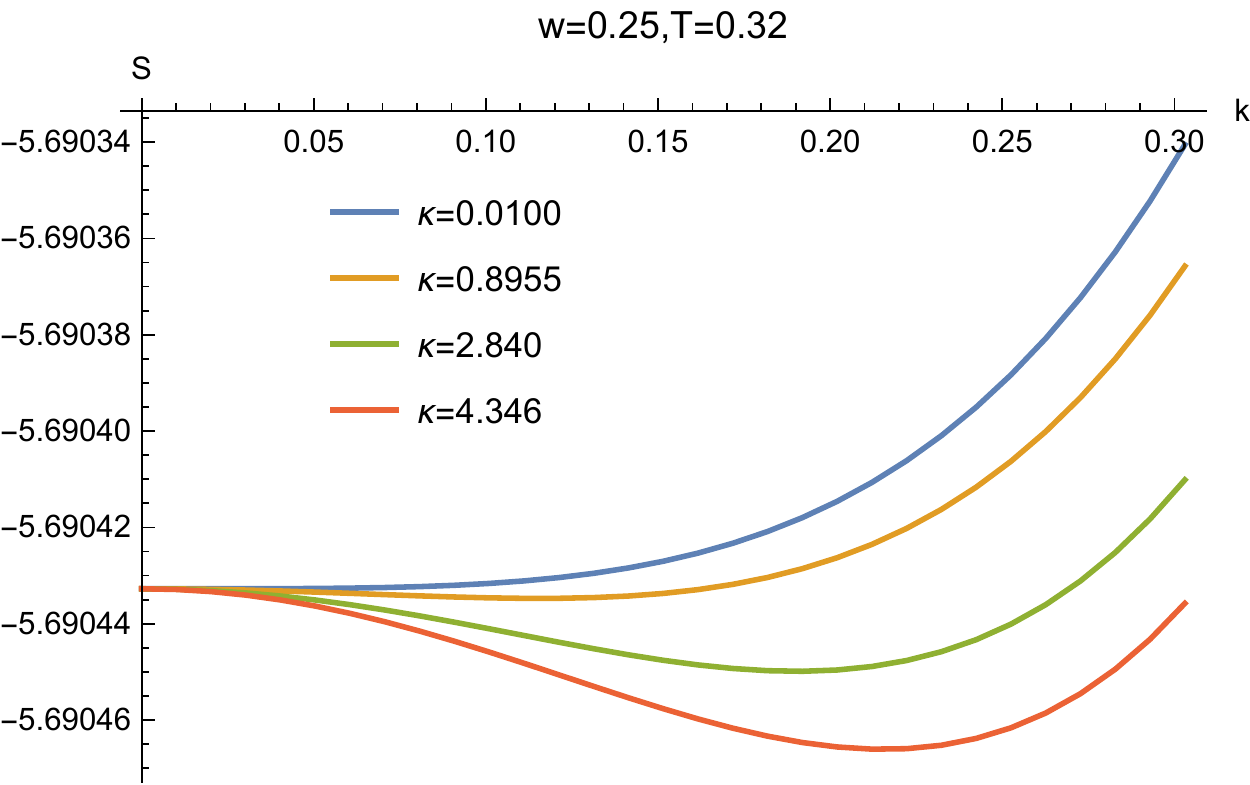}
	\caption{The relation between HEE and $k$ in the small configuration for model II. It can be seen that when $\kappa$ is very small, HEE increases monotonically with $k$, and with the increase of $\kappa$, the behavior of HEE decreases first and then increases with $k$.}
	\label{fig:case2_small_configuration_HEEk1}
\end{figure}

First, we study HEE vs $k$. We find that HEE increases monotonically with $k$ when $\kappa$ is relatively small. This can be deduced from the relationship between the $r_h$ and $k$ at relatively small values of $\kappa$. From the Hawking temperature \eqref{eq:hawkingtemperature2} and the horizon condition \eqref{eq:horizoncondition}, we can deduce that,
\begin{equation}\label{eq:fulltemp}
	T=-\frac{\frac{2 \kappa  k^2}{\left(\kappa  k^2 r_h+r_h^3\right) \left(\pi -2 \arctan\left(\frac{r_h}{\sqrt{\kappa } k}\right)\right){}^2}+\frac{k^4}{r_h^3}-6 r_h}{8 \pi }.
\end{equation}
Taking $\kappa\to 0$ limit, one obtain that,
\begin{equation}\label{eq:fulltempkappa0}
	T=\frac{6 r_h^4-k^4}{8 \pi  r_h^3}.
\end{equation}
Therefore, one find that,
\begin{equation}\label{eq:drhdksmallkappa}
	\partial_k r_h = \frac{4 k^3 r_h}{3 \left(2 r_h^4+k^4\right)}>0.
\end{equation}
The relationship \eqref{eq:drhdksmallkappa} suggests that $r_h$ monotonically increases with $k$ when $\kappa$ is small. The increasing $r_h$ means that the horizon moves closer to the boundary, and the minimum surface will stretch to the near-horizon region, and be affected by the thermal entropy. The fact that $r_h$ increases with $k$ also suggests that the thermal entropy increases with $k$. When $\kappa$ further increases, HEE decreases first and then increases with $k$. We show this result in Fig. \ref{fig:case2_small_configuration_HEEk1}.
The monotonically increasing behavior between the HEE and $k$ when $k$ is large enough cannot be analyzed by taking $\kappa \to 0$. Instead, when $k$ is relatively large, it is readily seen that $\, _2F_1\left(-\frac{1}{2},1;\frac{1}{2};-\frac{r_h^2}{\kappa  k^2}\right)$ will be small, then one can expand it as,
\begin{equation}\label{eq:drhdklargek}
	\begin{aligned}
		\partial_k r_h & = \frac{4 k r_h \left(2 \pi ^3 \kappa  k^4 r_h^2+\pi ^3 k^2 r_h^4-2 \kappa ^{3/2} k r_h^3+\pi  \kappa  r_h^4+\pi ^3 \kappa ^2 k^6\right)}{2 \kappa  k^2 r_h^2 \left(-4 \sqrt{\kappa } k r_h+3 \pi  r_h^2+\pi  \kappa  k^2\right)+3 \pi ^3 \left(2 r_h^4+k^4\right) \left(r_h^2+\kappa  k^2\right){}^2} \\
		               & = \frac{4 r_h}{3 k}+O\left(k ^{-2}\right) >0.
	\end{aligned}
\end{equation}
Therefore, when $k$ is relatively large, $r_h$ will monotonically increase with $k$. This explains the monotonically increasing behavior of HEE along $k$ at a large $k$.

\begin{figure}
	\centering
	\includegraphics[width = 0.6\textwidth]{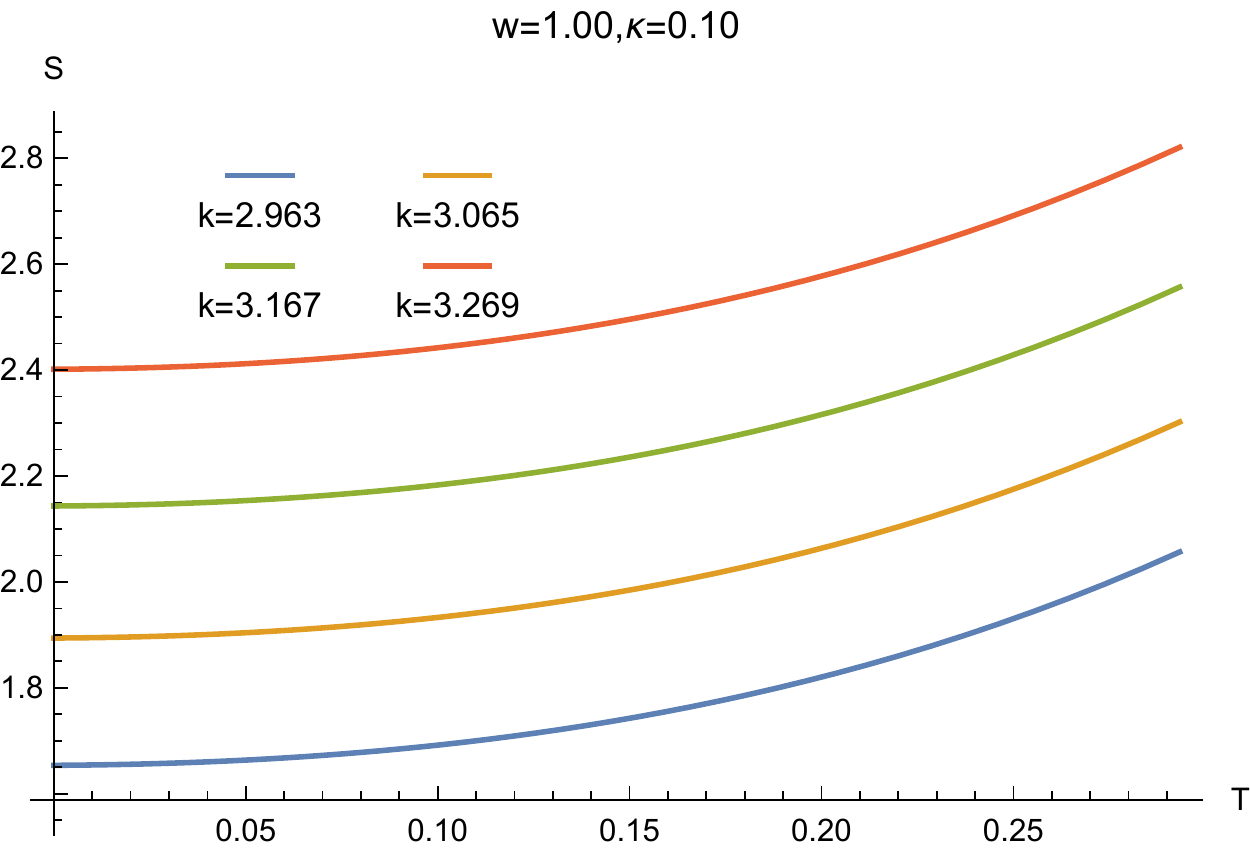}
	\caption{The relation between HEE and $T$ in the small configuration for model II. At other widths, HEE also increases monotonically with $T$.}
	\label{fig:case2_HEET1}
\end{figure}

Moreover, we also find that HEE increases monotonically with $T$, independent of the size of configuration and the values of $\kappa$, $k$ and $T$ (see Fig. \ref{fig:case2_HEET1}).
This phenomenon at high temperatures is understandable in an analytical manner. From \eqref{eq:fulltemp} one finds that $r_h$ is large when $T$ is large.
Varying the $r_h$ and we can find,
\begin{equation}\label{eq:drhdkv2}
	\partial_T r_h = \frac{8 \pi }{-\frac{8 \kappa ^{3/2} k^3}{J^3 r_h \left(r_h^2+\kappa  k^2\right){}^2}+\frac{2 \kappa  k^2 \left(3 r_h^2+\kappa  k^2\right)}{J^2 r_h^2 \left(r_h^2+\kappa  k^2\right){}^2}+\frac{3 k^4}{r_h^4}+6},
\end{equation}
where $J \equiv \pi -2 \arctan\left(\frac{r_h}{\sqrt{\kappa } k}\right)$. When $r_h$ increases with $T$ and becomes relatively large, $J$ can be expanded as,
\begin{equation}\label{eq:expandA}
	J = \frac{\pi }{2}-\frac{\sqrt{\kappa } k}{r_h} + O(r_h^{-2}).
\end{equation}
Therefore, we find,
\begin{equation}\label{eq:drhdtv1}
	\begin{aligned}
		\partial_T r_h & = \frac{16 \pi  r_h^4 \left(r_h^2+\kappa  k^2\right)}{r_h^4 \left(12 \kappa  k^2+1\right)+6 k^4 r_h^2+12 r_h^6+6 \kappa  k^6} \\
		               & =\frac{4 \pi }{3}+O\left(1/r_h \right) > 0.
	\end{aligned}
\end{equation}
From this relation, we find that HEE will increase with $T$. Also, when the temperature is relatively large, the slope of $r_h$ vs $T$ will be close to $4\pi/3$.

\begin{figure}
	\centering
	\includegraphics[width = 0.49\textwidth]{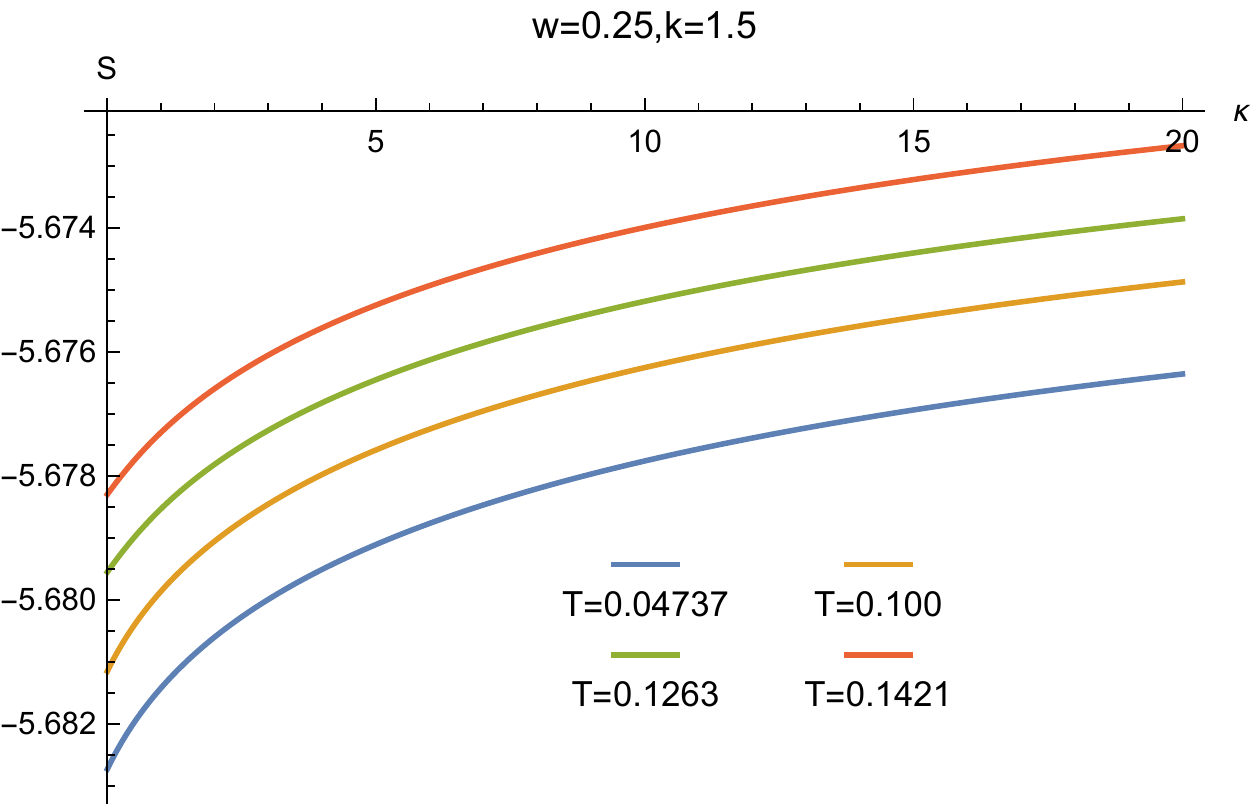}
	\includegraphics[width = 0.49\textwidth]{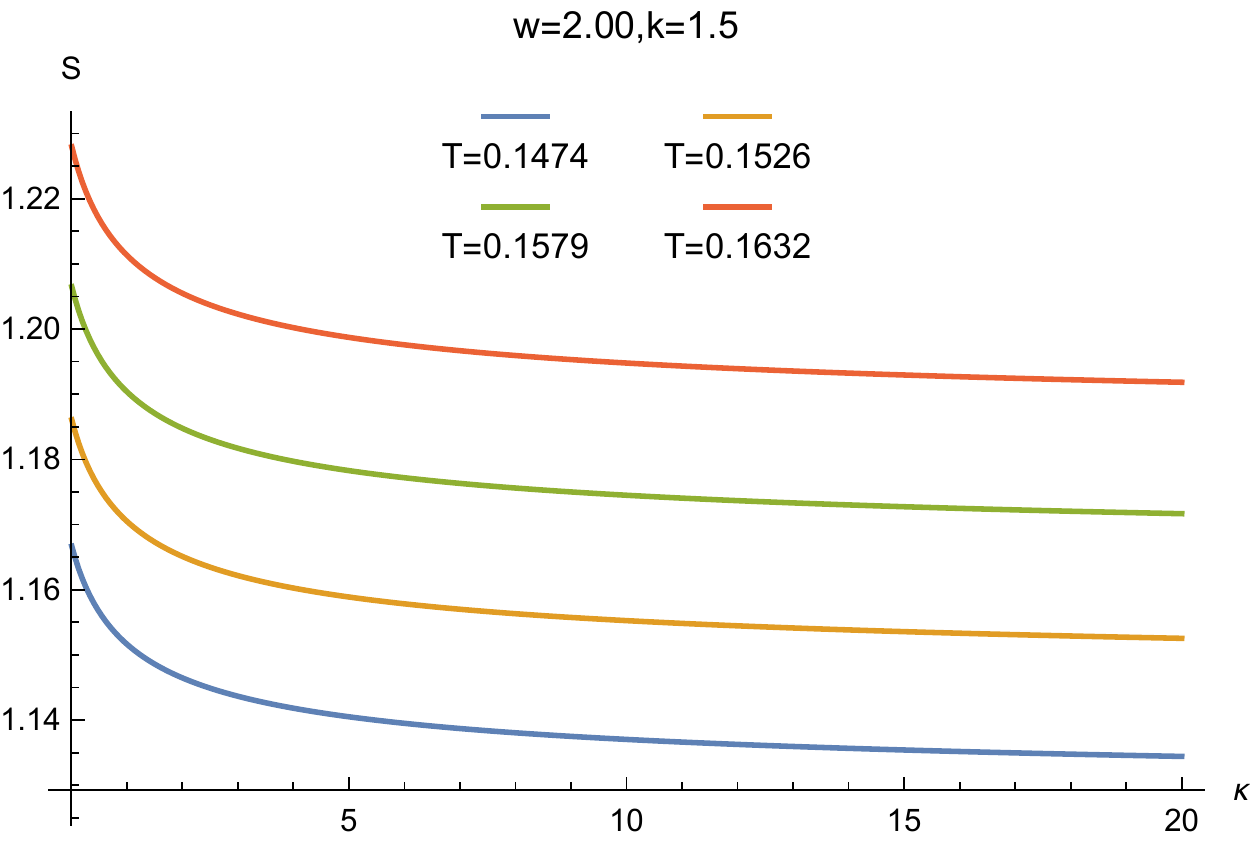}
	\caption{The left and right graphs show the relationship between HEE and $\kappa$ in the small and large configurations for model II respectively. It can be seen that HEE monotonically increases with $\kappa$ in small configurations and decreases monotonically with $\kappa$ in large configurations.}
	\label{fig:case2_HEEkappa1}
\end{figure}

Next, we discuss the relationship between HEE and $\kappa$. HEE increases monotonically with $\kappa$ for small configurations (see the left plot of Fig. \ref{fig:case2_HEEkappa1}); while for large configurations, HEE decreases monotonically with $\kappa$ (see the right plot of Fig. \ref{fig:case2_HEEkappa1}).
For large configurations, the decreasing HEE along $\kappa$ allows an analytical understanding when $\kappa$ is small or large enough. First, when $\kappa$ is small enough, we will find that,
\begin{equation}\label{eq:drhdkappav2}
	\partial_\kappa r_h = -\frac{k^2 r_h}{3 \left(12 r_h^4+r_h^2+6 k^4\right)}+O\left(\kappa ^1\right) <0.
\end{equation}
Therefore, $r_h$ will decrease with the increase of $\kappa$, and HEE will also decrease with the increase of $\kappa$. Second, when $\kappa$ is large, one finds that,
\begin{equation}\label{eq:drhdkkappv3}
	\partial_\kappa r_h = -\frac{r_h^5}{6 \kappa ^2 k^2 \left(2 r_h^4+k^4\right)}+O\left(\kappa ^{-5/2}\right) <0.
\end{equation}
Therefore, again, we find that $r_h$ decreases with $\kappa$ when $\kappa$ is large. Therefore, $r_h$ will decrease with the increase of $\kappa$, and HEE will again decrease with the increase of $\kappa$.

\begin{figure}
	\centering
	\includegraphics[width = 0.6\textwidth]{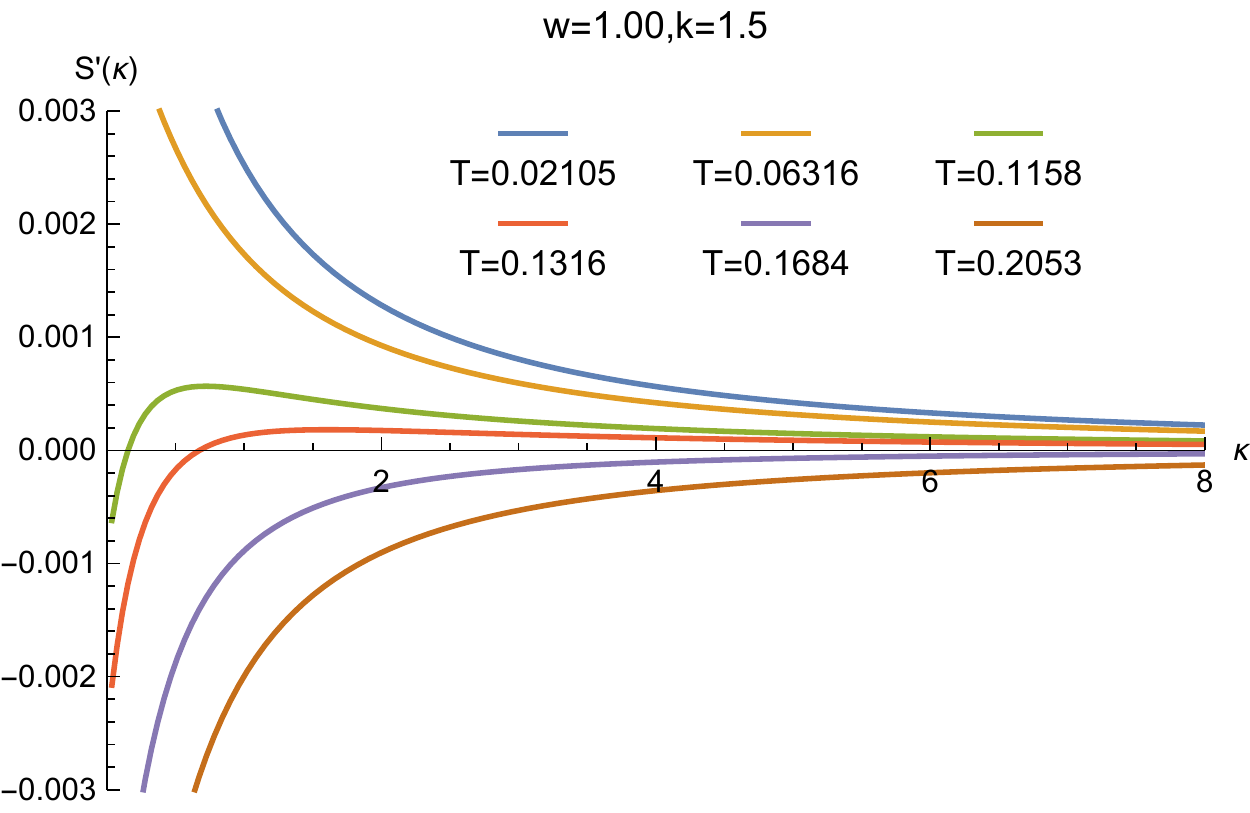}
	\caption{The relationship between $\frac{ \partial S }{ \partial \kappa}$ and $\kappa$ in one of the intermediate configurations for model II. It can be seen that at low temperature, HEE monotonically increased with $\kappa$, and with the increase of temperature, the behavior of HEE with $\kappa$ decreased first and then increased, and with the further increase of temperature, HEE monotonically decreased with $\kappa$.}
	\label{fig:case2_HEEkappa2}
\end{figure}

In the case of intermediate configuration, the behavior of HEE with $\kappa$ becomes complicated, which is related to the detailed values of the widths, $k$ and $T$. The derivative of HEE with ${\kappa}$ (Fig. \ref{fig:case2_HEEkappa2}) can clearly show the phenomena of intermediate configuration, which we expand in three cases.
\begin{enumerate}
	\item When $T$ is small, $S'(\kappa)$ is always positive, indicating that HEE increases monotonically with $\kappa$. This is consistent with the phenomenon in the left plot of Fig. \ref{fig:case2_HEEkappa1}.
	\item When $T$ gradually increases to a temperature around $T = 0.1158$, $S'(\kappa)<0$ when $\kappa$ is very small, but increases to be positive when $\kappa$ is large. This indicates that HEE first decreases and then increases with $\kappa$.
	\item When $T$ increases further, $S'(\kappa)$ is always negative, indicating that HEE decreases monotonically with $\kappa$. This is consistent with the phenomenon in the right plot of Fig. \ref{fig:case2_HEEkappa1}. In addition, a physical understanding is allowed here. With the increase of temperature, $r_h$ will also increase. Therefore, the minimum surface of the intermediate configurations will also be close to the horizon of the black brane, so the relationship between HEE and $\kappa$ will be the same as that of the large configuration.
\end{enumerate}

\subsection{Mutual information}

For the configurations with nontrivial MI,
\begin{equation}
	I(A;C) = S(A) + S(C) - S (B) - S(A \cup B \cup C).
\end{equation}
Consequently, the derivative of MI with any system parameter $x$ will be,
\begin{equation}
	\frac{ \partial I(A;C) }{ \partial x }  = \frac{ \partial S(A) }{ \partial x } + \frac{ \partial S(C) }{ \partial x } - \frac{ \partial S(B) }{ \partial x } - \frac{ \partial S(A \cup B \cup C) }{ \partial x }.
\end{equation}
We find that $ \frac{ \partial S(B) }{ \partial x } + \frac{ \partial S(A \cup B \cup C) }{ \partial x } $ is larger than $ \frac{ \partial S(A) }{ \partial x } + \frac{ \partial S(C) }{ \partial x } $ in most cases. In other words, the change of the MI with parameters is mainly contributed by the part closer to horizon. The reason is that the region closer to horizon is farther away from the boundary, so it is more affected by the deviation of AdS. Therefore, the change of MI with parameters will be completely opposite to that of HEE. Next we show numerical evidences.

\begin{figure}
	\centering
	\includegraphics[width = 0.49\textwidth]{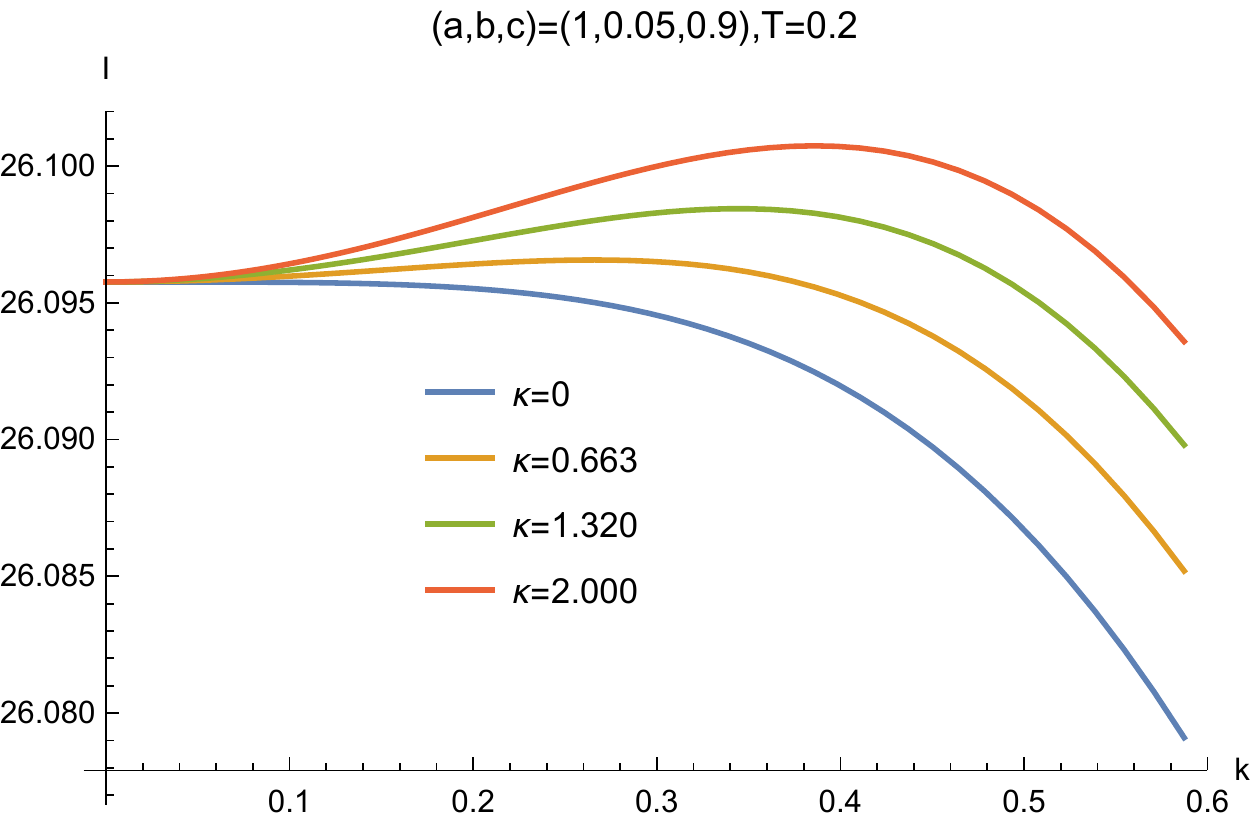}
	\includegraphics[width = 0.49\textwidth]{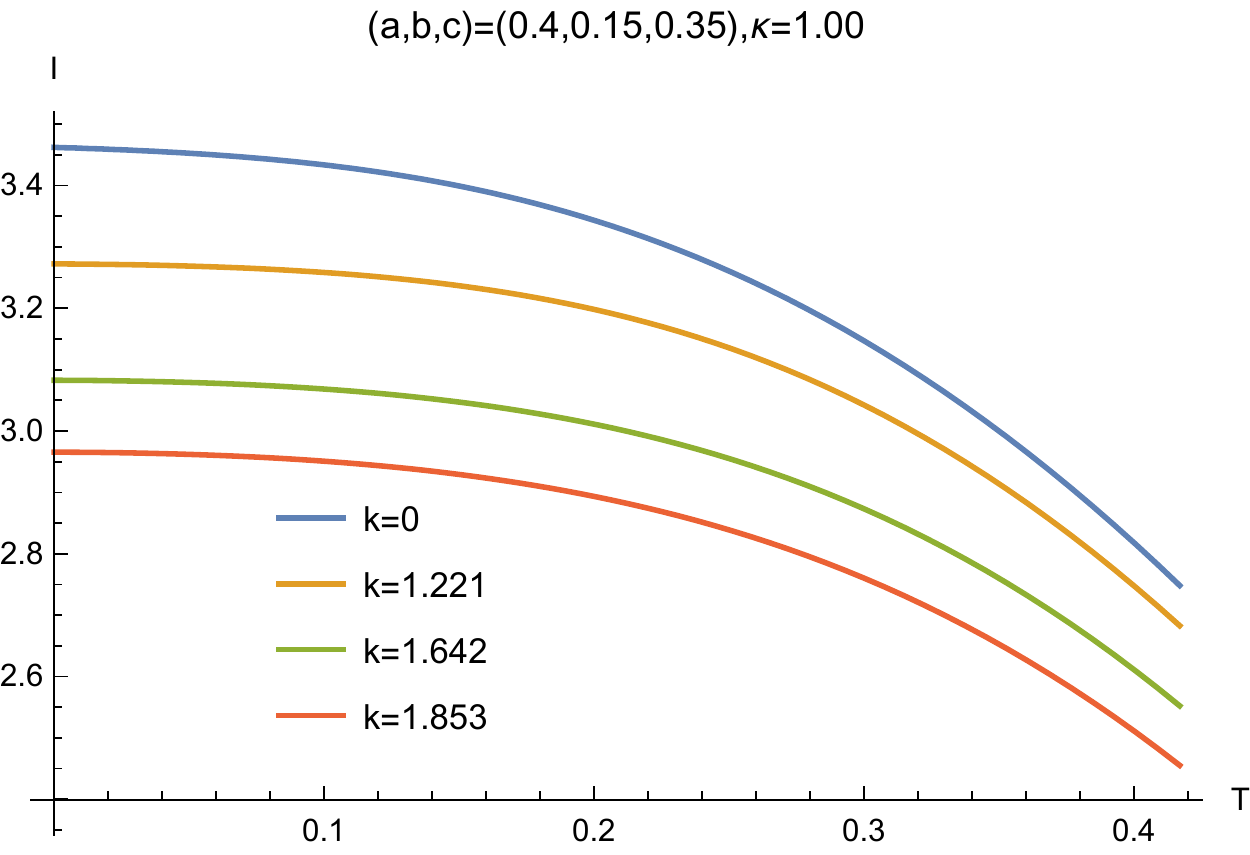}
	\caption{
		The figure on the left shows the relationship between MI and $k$ for model II. It can be seen that when $\kappa$ is very small, MI decreases monotonically with $k$, and with the increase of $\kappa$, the behavior of MI increases first and then decreases with $k$. The figure on the right shows the relationship between MI and $T$ for model II. It can be seen that MI decreases monotonically with $T$.
	}
	\label{fig:case2_MIkT}
\end{figure}

MI decreases monotonically with $k$ when the configuration and $\kappa$ are not large. With the increase of configuration or $\kappa$, MI increases first and then decreases. We show this result as the left plot in Fig. \ref{fig:case2_MIkT}. It is almost the opposite of the behavior shown in Fig. \ref{fig:case2_small_configuration_HEEk1}.
Also, we find that MI decreases monotonically with $T$ (see the right plot of Fig. \ref{fig:case2_MIkT}), which is independent of the size of configuration and the values of $\kappa$, $k$ and $T$. This is again opposite to HEE shown in Fig. \ref{fig:case2_HEET1}. Finally, it can be inferred from the complicated relationship between HEE and $\kappa$ that the relationship between MI and $\kappa$ will also show a complicated but opposite relationship to HEE.

Next, we discuss the relationship between EWCS and system parameters in detail.

\subsection{Entanglement Wedge Minimum Cross Section}

First, we discuss the relationship between EWCS and $k$. The relationship between EWCS and $k$ is related to the specific configuration, which we divide into the large configuration case and the small configuration case.

\begin{enumerate}
	\item For large configurations, EWCS increases monotonically with $k$, independent of the values of $T$ and $\kappa$. We show this result in Fig. \ref{fig:case2_large_configuration_EoPk1}, and we also get the same conclusion in other large configurations.
	      \begin{figure}
		      \centering
		      \includegraphics[width = 0.6\textwidth]{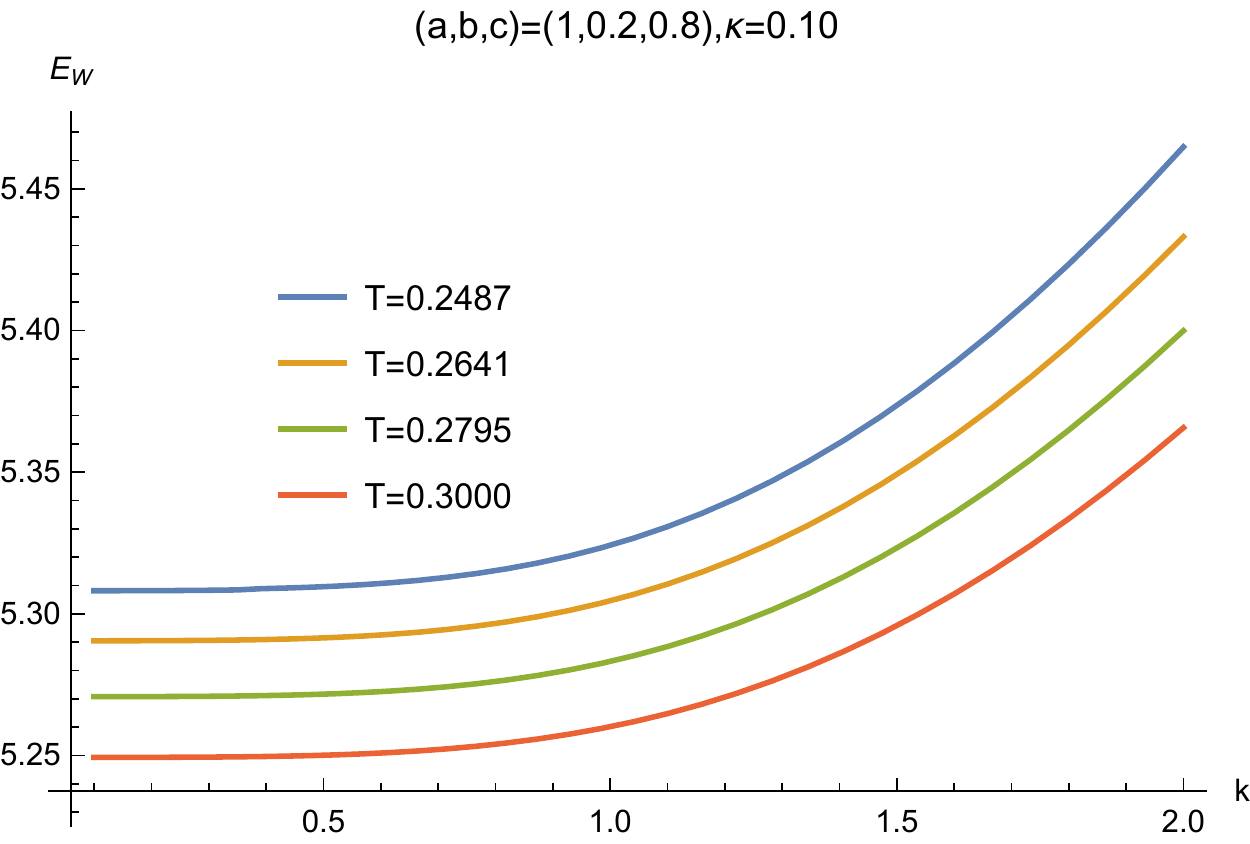}
		      \caption{The relation between EWCS and $k$ in the large configuration for model II. It can be seen that EWCS increases monotonically with $k$, and has the same monotonic behavior with other large configurations and the values of $T$ and $\kappa$.}
		      \label{fig:case2_large_configuration_EoPk1}
	      \end{figure}
	\item For small configuration, EWCS exhibits rich phenomena with $k$. When the values of $T$ and $\kappa$ are appropriate, EWCS will decrease first and then increase with $k$ (see Fig. \ref{fig:case2smalleopk1}).
	      \begin{figure}
		      \centering
		      \includegraphics[width = 0.6\textwidth]{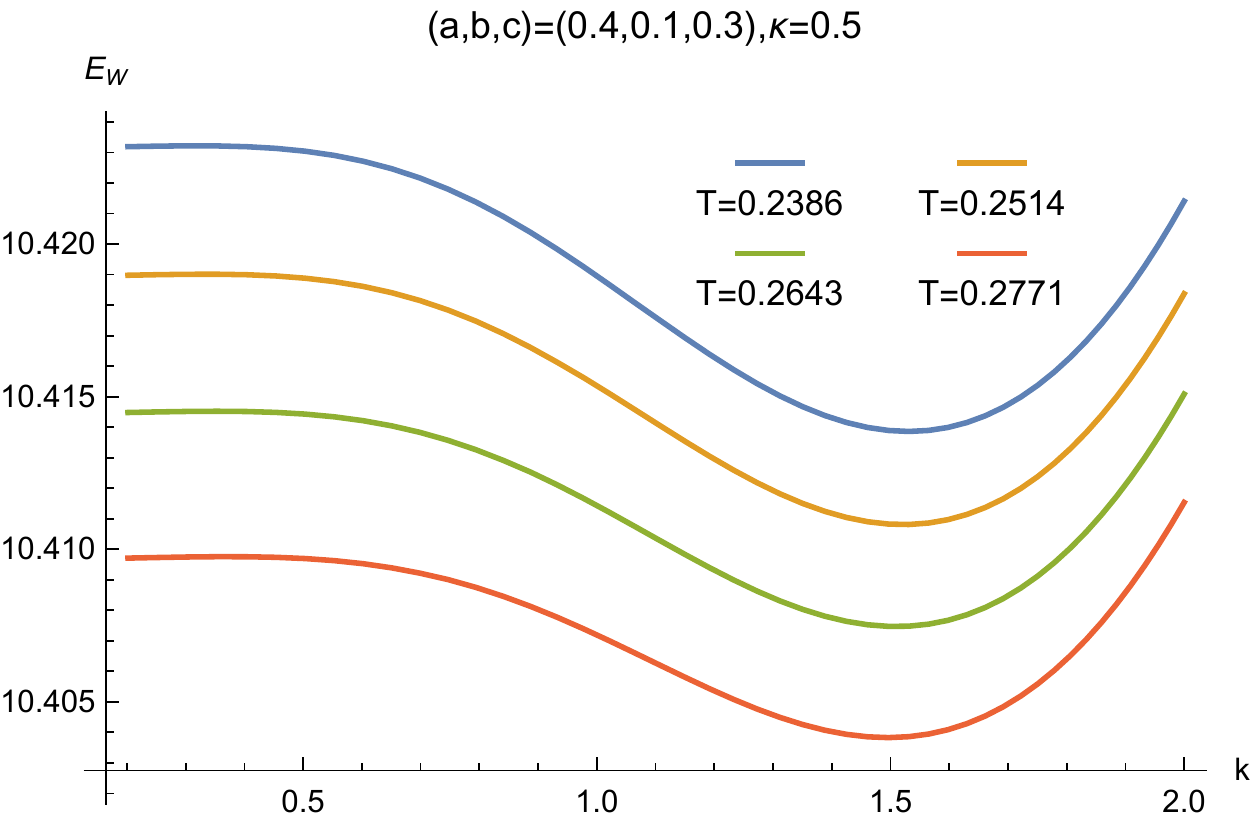}
		      \caption{The relation between EWCS and $k$ in the small configuration for model II. It can be seen that when the values of $T$ and $\kappa$ are appropriate, EWCS first decreases and then increases with $k$.}
		      \label{fig:case2smalleopk1}
	      \end{figure}
	      However, with the increase of $T$ or $\kappa$, the decreasing behavior of EWCS with $k$ becomes less significant, even disappears, and becomes monotonically increasing with $k$ (see Fig. \ref{fig:case2_small_configuration_EoPk2}).
	      \begin{figure}
		      \centering
		      \includegraphics[width = 0.45\textwidth]{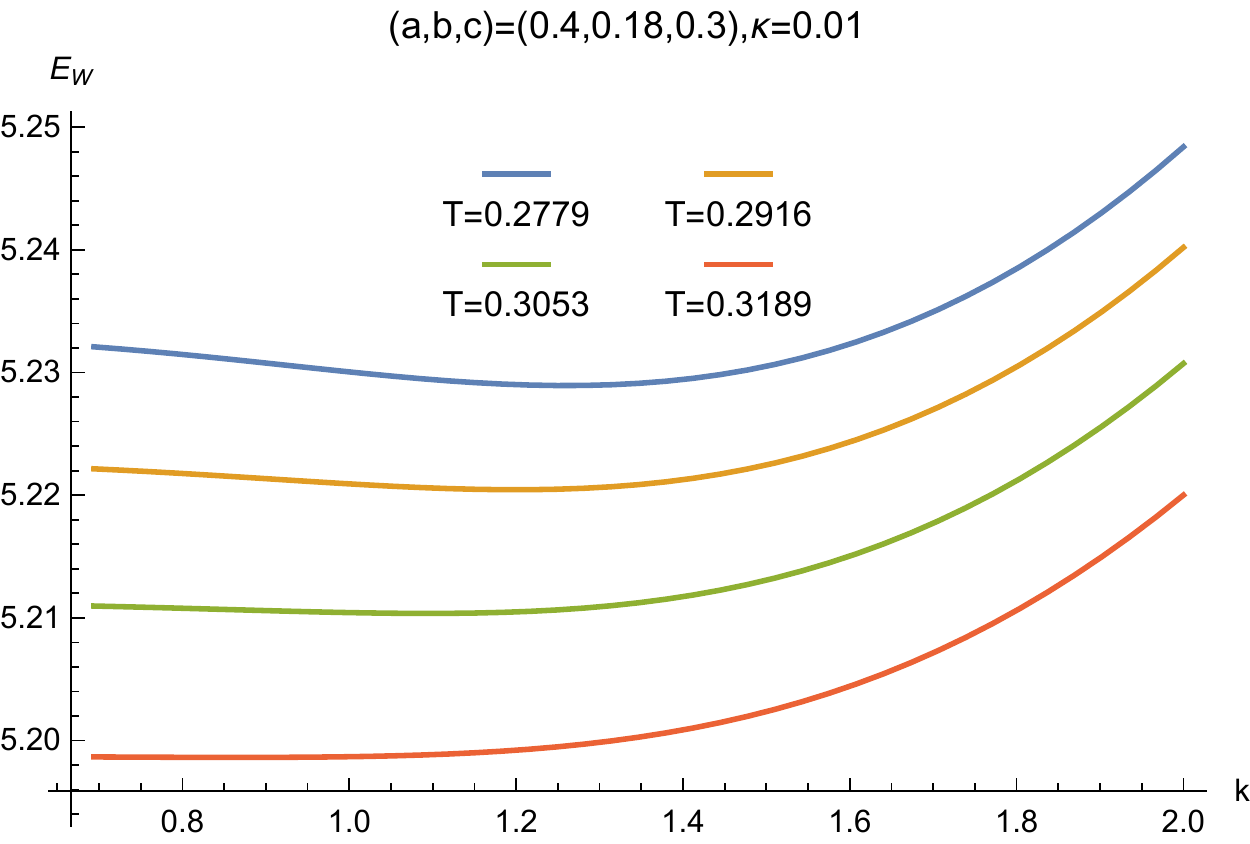}
		      \includegraphics[width = 0.45\textwidth]{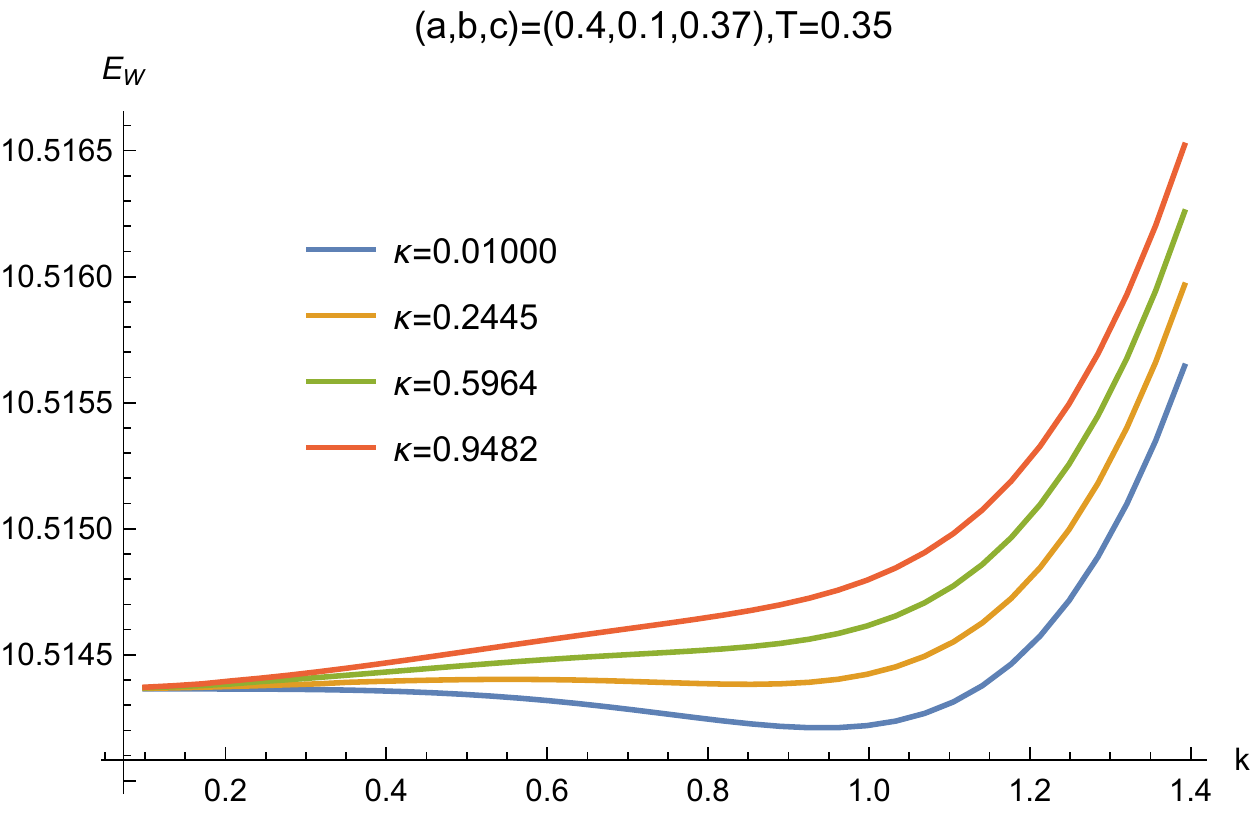}
		      \caption{
			      The left and right figures show the relation between EWCS and $k$ under different small configurations, $T$ and $\kappa$ for model II. It can be seen from the left figure that when $T$ increases, the decreasing behavior of EWCS with $k$ weakens or even disappears. It can be seen from the right figure that when $\kappa$ increases, the decreasing behavior of EWCS with $k$ weakens or even disappears.
		      }
		      \label{fig:case2_small_configuration_EoPk2}
	      \end{figure}
\end{enumerate}

The above phenomena show that EWCS presents rich phenomena with $k$. It is particularly noteworthy that when $k$ is large, EWCS always increases with the increase of $k$. Another obvious universal behavior is that when $k$ is small, EWCS shows a flat behavior with $k$. These two common phenomena can be understood through an analytical treatment. Note that the EWCS is given by,
\begin{equation}\label{eq:ewcsform}
	\begin{aligned}
		E_W & = \int \sqrt{g_{yy}} \sqrt{g_{xx} dx^2 + g_{rr} dr^2} \\
		    & = \int r \sqrt{r^2 dx^2 + \frac{dr^2}{f(r)}}.
	\end{aligned}
\end{equation}
Therefore, the $E_W$ behavior with parameters depends on the $f(r)$.
\begin{equation}\label{eq:ewcsd}
	\partial_\eta E_W =  - \int \frac{ r \partial_\eta f}{2 f^2 \sqrt{\frac{dr^2}{f}+dx^2 r^2}}dr^2.
\end{equation}
In large $k$-limit, we will find that
\begin{equation}\label{eq:2f1limit}
	\, _2F_1\left(-\frac{1}{2},1;\frac{1}{2};-\frac{r_h^2}{k^2 \kappa }\right)\to 1.
\end{equation}
By working out $ \partial_{k} r_h $ from \eqref{eq:hawkingtemperature2}, we can obtain that,
\begin{equation}\label{eq:dfdk}
	\partial_k f = \frac{2 k^3 \left(3 r_h-2 r\right)}{3 r^2 r_h}+O\left(1/k\right).
\end{equation}
Note that, the separation size $b$ cannot be too large, otherwise the EWCS vanishes following from a vanishing MI. The $E_W$ is mainly contributed from the region away from the near horizon region, i.e., $r\gg r_h$. Therefore, the main contribution to \eqref{eq:ewcsd} from \eqref{eq:dfdk} will be positive.
This relation explains the universal monotonically increasing behavior revealt in Fig. \ref{fig:case2_large_configuration_EoPk1}, \ref{fig:case2smalleopk1} and \ref{fig:case2_small_configuration_EoPk2}. For relatively small $k$, one can also expand the $\partial_k f$ as,
\begin{equation}\label{eq:dewdfsmallk}
	\partial_k f = -\frac{\kappa  k \left(-2 r^2 r_h+r_h^3+r^3\right)}{4 r^4 r_h}+O\left(k^2\right).
\end{equation}
It means that $\partial_k f \sim k$ for small $k$, which explains the flat behavior for $E_W$. However, whether the $E_W$ increases or decreases with $k$ depends on the specific configuration in consideration.

\begin{figure}
	\centering
	\includegraphics[width = 0.6\textwidth]{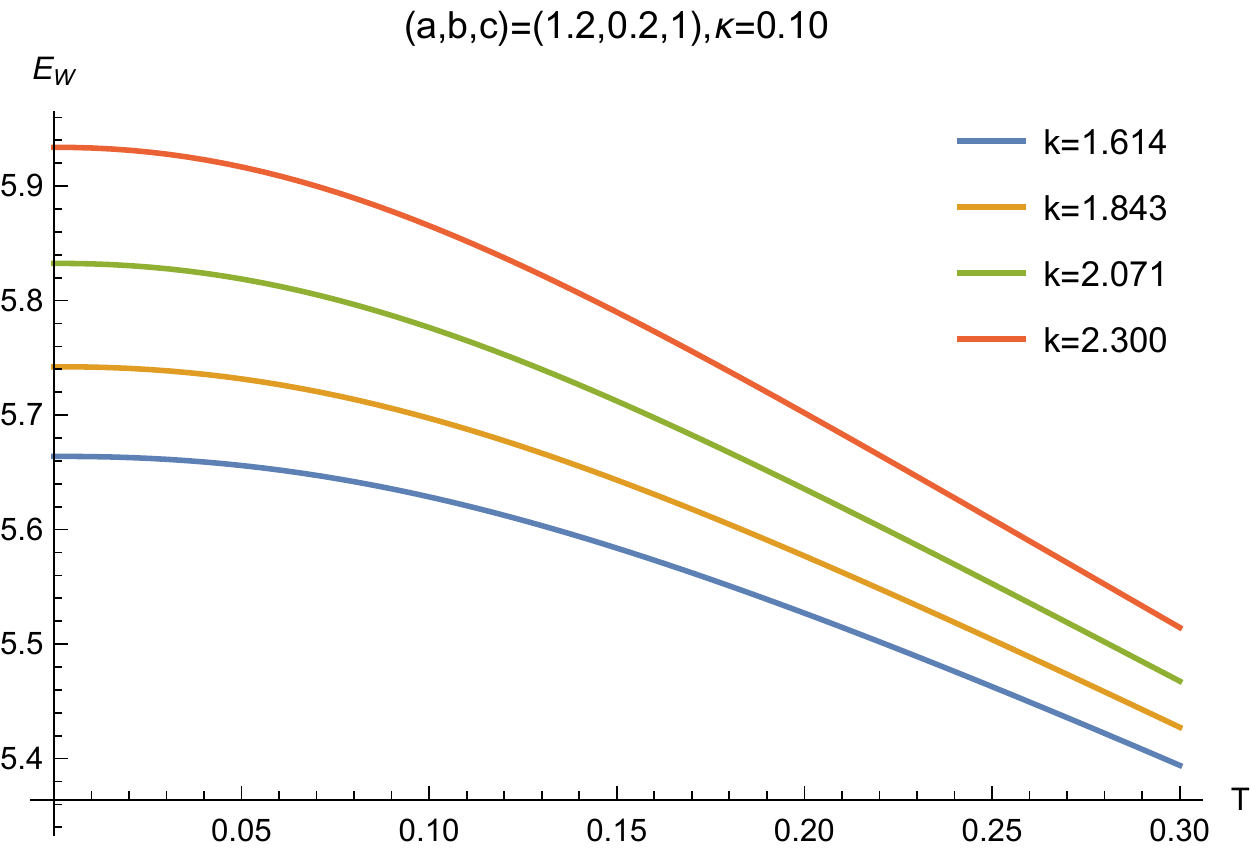}
	\caption{The relation between EWCS and $T$ for model II. It can be seen from the figure that EWCS decreases monotonically with $T$, and it also has the same monotonic behavior under other configurations, $k$ and $\kappa$.}
	\label{fig:case2eopt1}
\end{figure}

For the relationship between EWCS and $T$, we find that EWCS decreases monotonically with $T$, independent of the configuration, $\kappa$ and $k$. We show this result in Fig. \ref{fig:case2eopt1}. This can be understood from the physical point of view that increasing the temperature will destroy the entanglement between the subregions. From the theory of quantum statistics, it can be understood that when the temperature of the system increases, the density matrix of the subsystem will be close to a product state of which each one is defined by the thermal density matrix. From the perspective of holographic duality, we can get a more vivid understanding. When the temperature increases, the minimum surface will approach the horizon of the black brane, so the HEE will be gradually controlled by the thermal entropy. When the temperature is very high, all the minimum surfaces will be contributed by the event horizon of the black brane, and MI is zero. Consequently, EWCS will vanish.

\begin{figure}
	\centering
	\includegraphics[width = 0.6\textwidth]{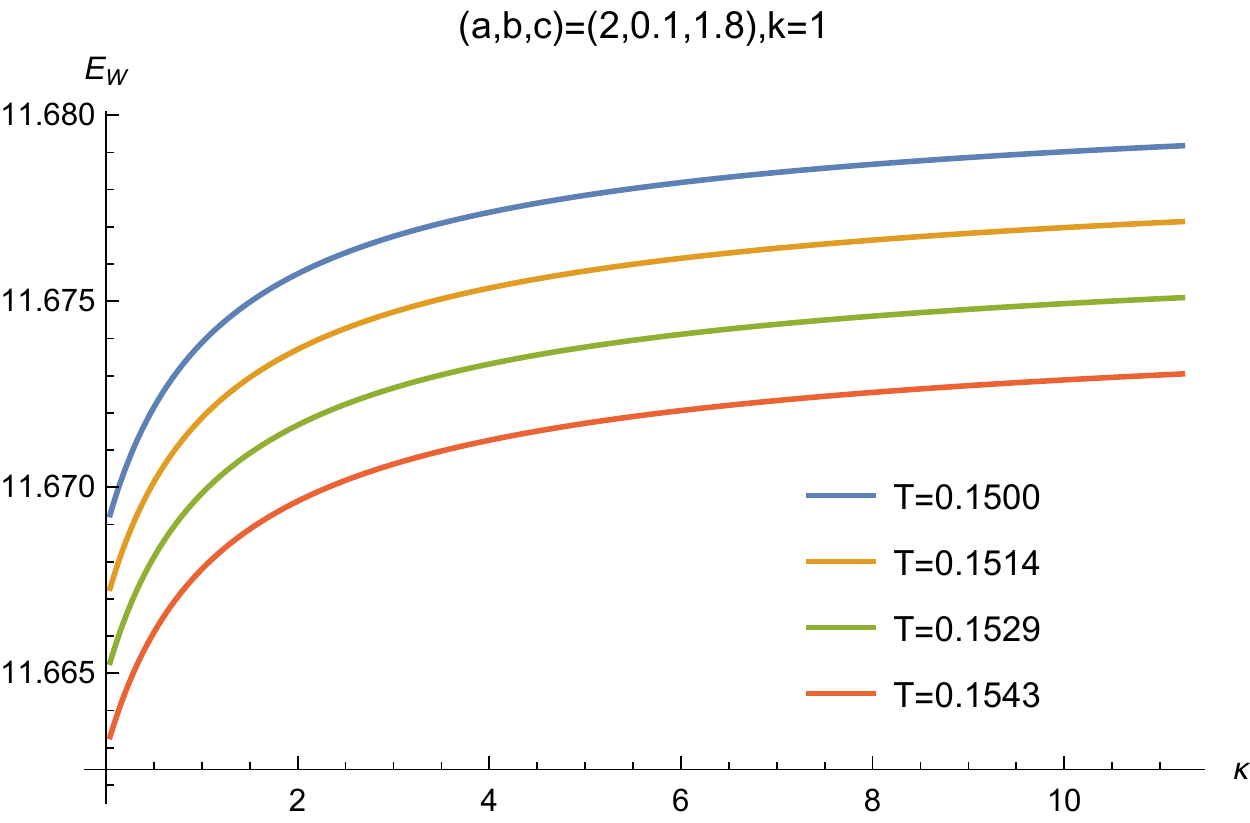}
	\caption{The relation between EWCS and $\kappa$ in model II. It can be seen from the figure that EWCS increases monotonically with $\kappa$, and it also has the same monotonic behavior under other configurations, $k$ and $T$.}
	\label{fig:case2_EoPkappa1}
\end{figure}

For the relationship between EWCS and $\kappa$, we find that EWCS increases monotonically with $\kappa$, and this behavior is independent of configuration, $T$ and $k$. We show this result in Fig. \ref{fig:case2_EoPkappa1}. Further, it is noted that both HEE and MI showed non-monotonic behavior with $\kappa$. This shows that EWCS does reflect the different properties of mixed state entanglement of the system.
When $\kappa$ is large, $E_W$ increases monotonically with $\kappa$ and it becomes more and more flat. This flat behavior can also be analyzed analytically. Expanding $\partial_\kappa f$ around $1/\kappa \to 0$ gives,
\begin{equation}\label{eq:dewdkappa}
	\partial_\kappa f = \frac{r-r_h}{2 \pi ^2 \kappa  r}+O\left(\kappa^{-3/2}\right).
\end{equation}
Therefore, the $\partial_\kappa f$ becomes flat, which explains the flat behavior of $E_W$ along $\kappa$.

\section{Discussion}
\label{sec:discuss}

In this paper, we study the behavior of EWCS with system parameters of linear axion model I, and we also study the behavior of HEE and MI with three important parameters in the linear axion model II. In the exploration of model I, we find that the EWCS of asymmetric configuration can show obvious non-monotonicity, which is an important supplement to the monotonicity of EWCS of previous symmetric configuration \cite{Liu:2020blk}.

In model II, we find that the behavior of HEE and MI with three parameters are always opposite, which reflects the tight connection between HEE and MI. Moreover, we also found that HEE, MI and EWCS show monotonic behavior with temperature. In addition, we find that these information-related physical quantities show richer non-monotonic behavior with the linear axion parameter $k$. A more important phenomenon is that HEE and MI show non-monotonic behavior with the coupling constant $\kappa$, but EWCS shows monotonic behavior. These phenomena show that the mixed state entanglement EWCS captured is indeed distinct from the HEE and MI, those suffer from thermal entropy pollution. We also provide analytical and physical understanding of the above phenomena obtained from numerics.

Next, we point out several directions worthy of further study. First, the difference between EWCS and HEE and MI is worth studying in more general cases. For example, when $V(X) \neq X^2$, the conclusions in this paper are worth further testing and discussion. Second, we can further discuss the relationship between the measurement of entanglement of other types of mixed states and EWCS in this paper, such as negativity \cite{Kudler-Flam:2018qjo}. Finally, the analytical understanding in this paper can be further extended to more general parameter ranges, which will provide analytical evidence for the rich numerical results in this paper.

\section*{Acknowledgments}

Peng Liu would like to thank Yun-Ha Zha for her kind encouragement during this work. This work is supported by the Natural Science Foundation of China under Grant No. 11805083, 11905083, 12005077 and Guangdong Basic and Applied Basic Research Foundation (2021A1515012374).

\end{document}